\pdfoutput=1
\documentclass[aps,prc,reprint,superscriptaddress,amsmath,amssymb,floatfix]{revtex4-2}
\usepackage{graphicx}
\usepackage{xcolor}
\usepackage{bm}
\usepackage[colorlinks=true,linkcolor=blue,citecolor=blue,urlcolor=blue]{hyperref}

\begin{document}

\title{Formation of superheavy elements 119 and 120 in the $^{50}$Ti+$^{249}$Bk and $^{50}$Ti+$^{249}$Cf reactions: dinuclear system dynamics, survival probability, and decay chains of the evaporation residues}

\author{Bakhodir Kayumov}
\email{b.kayumov@newuu.uz}
\affiliation{Department of Physics, New Uzbekistan University, Tashkent 100000, Uzbekistan}
\affiliation{Laboratory of Theoretical Nuclear Physics, Institute of Nuclear Physics, Tashkent 100214, Uzbekistan}

\date{\today}

\begin{abstract}
The synthesis of elements 119 and 120 is one of the main goals of the present experimental programs with $^{50}$Ti beams. In this work, the $^{50}$Ti+$^{249}$Bk and $^{50}$Ti+$^{249}$Cf reactions leading to the compound nuclei $^{299}$119 and $^{299}$120 are studied in the framework of the dinuclear system (DNS) model. The capture cross section has been calculated by solving the dynamical equations of motion for all orientation angles of the deformed target nucleus and for the zero-point vibrations of the projectile. The charge distribution of the DNS fragments has been obtained from the transport master equation with transition rates obeying detailed balance, and the compound nucleus formation probability has been calculated as the branching ratio of the level densities over the intrinsic fusion, quasifission, and symmetry barriers. The survival probability has been calculated in the Hauser-Feshbach formalism for every excitation energy and angular momentum of the compound nucleus, with the fission barriers of the macroscopic-microscopic model with triaxial saddle points. The shell correction is damped with the excitation energy through the level density parameters at the ground state and at the saddle point, and with the angular momentum through one factor applied to the whole shell contribution; the normalization of the saddle-point level density has been fixed by the measured evaporation residue (ER) cross sections of the $^{48}$Ca+$^{244}$Pu reaction. The largest ER cross sections are obtained in the 3n channel, 18 fb at $E^*_{\rm CN}=32$ MeV for $^{50}$Ti+$^{249}$Cf and 27 fb at 33 MeV for $^{50}$Ti+$^{249}$Bk. At the energy of the GSI experiments, the calculated sums of the 3n and 4n cross sections, 4 and 6 fb, respectively, are well below the upper limits of 200 and 65 fb obtained there. With the normalization refitted, the alternative mass tables and fission barriers considered here give the maximum 3n cross sections smaller by up to a factor of 25. The $\alpha$-decay chains of the residues have been predicted; the isotopes of element 120 are expected to have half-lives of 1--12 $\mu$s and those of element 119 of 16--230 $\mu$s.
\end{abstract}

\maketitle

\section{INTRODUCTION}
\label{sec:intro}

The synthesis of superheavy elements (SHEs) is a crucial area of research in nuclear physics, providing insights into the stability and structure of nuclei at the extreme limits of the periodic table \cite{LRP2023,Smits2024,Giuliani2019,Godbey2025}. The elements with $Z=114$--118 have been synthesized in hot fusion reactions with $^{48}$Ca beams on actinide targets. Since targets heavier than californium are not available in sufficient amounts, the synthesis of elements 119 and 120 requires heavier projectiles, such as $^{50}$Ti, $^{51}$V, and $^{54}$Cr \cite{Zagrebaev2008,Adamian2020,Kayumov2022,Kayumov2025}. The cross sections expected for these elements are at the level of tens of femtobarns or lower \cite{Zagrebaev2008,Adamian2020}. The $^{50}$Ti+$^{249}$Bk and $^{50}$Ti+$^{249}$Cf reactions have been studied at the gas-filled separator TASCA at GSI, and no decay chain was observed at the cross-section sensitivity levels of 65 and 200 fb, respectively \cite{Khuyagbaatar2020}. On the other hand, the production of livermorium in the $^{50}$Ti+$^{244}$Pu reaction at LBNL \cite{Gates2024} and in the $^{50}$Ti+$^{242}$Pu reaction at the JINR Superheavy Element Factory \cite{Oganessian2025}, with cross sections of 220--440 fb, has demonstrated that the reactions with $^{50}$Ti are about one order of magnitude more favorable than those with $^{54}$Cr \cite{Oganessian2025}.

In our previous work \cite{Kayumov2025}, the $^{50}$Ti+$^{244}$Pu and $^{50}$Ti+$^{251}$Cf reactions were studied in the dinuclear system (DNS) model, and the calculated 4n cross section for $^{50}$Ti+$^{244}$Pu was found in good agreement with the LBNL measurement. In the present work, the $^{50}$Ti+$^{249}$Bk and $^{50}$Ti+$^{249}$Cf reactions are studied with several refinements of the method. First, the rotation of the DNS is treated in the sticking limit with the total angular momentum conserved, $L=L_0$, so that the driving potential does not depend on the collision energy. Second, the evolution of the charge distribution of the DNS fragments is calculated with the transition rates obeying detailed balance with respect to the level density of the DNS, and the compound nucleus formation probability is obtained from this distribution as the branching ratio of the level densities over the fusion, quasifission, and symmetry barriers \cite{Kayumov2022,NasirovKayumov2024}. Third, the survival probability is calculated in the Hauser-Feshbach (HF) formalism \cite{HauserFeshbach1952}, in which the population of the excited nuclei is followed in excitation energy and spin along the evaporation cascade, separately for each excitation energy and angular momentum of the compound nucleus; this makes it possible to determine the angular momentum window of the evaporation residues (ERs). In the fission width, the shell correction is damped only once: with the excitation energy through the level density parameters at the ground state and at the saddle point, and with the angular momentum through one factor applied to the whole shell contribution to the barrier and to the level densities. The fission barriers and the saddle-point shapes are taken from the macroscopic-microscopic calculation with triaxial saddle points \cite{Jachimowicz2021}, and the normalization of the saddle-point level density, as in the EMPIRE code \cite{EMPIRE32}, is fixed by the measured ER cross sections of the $^{48}$Ca+$^{244}$Pu reaction. The results are compared with those of the spin-independent cascade method \cite{KayumovAmirov}, in which the energy distribution of the excited nuclei is propagated through the evaporation steps without the spin degree of freedom and the neutron-fission competition is described by the Vandenbosch-Huizenga expression \cite{Vandenbosch1973}. Finally, the laboratory angular distribution of the ERs and the decay chains of the ER nuclei are calculated, since these quantities are needed for the identification of the products in an experiment.

This paper is structured as follows: Sec.~\ref{sec:capture} discusses the capture cross section calculations, Sec.~\ref{sec:cn} presents the compound nucleus formation probabilities, Sec.~\ref{sec:survival} describes the survival probability, and Sec.~\ref{sec:er} analyzes the ER cross sections and the angular distribution of the residues. The decay chains of the ERs are discussed in Sec.~\ref{sec:chains}. Finally, the conclusions are reported in Sec.~\ref{sec:concl}. The full table of the predicted decay chains is given in the Appendix.

\section{CAPTURE CROSS SECTION}
\label{sec:capture}

The capture of the projectile by the target nucleus is calculated by solving the equations of motion for the relative distance $R$ and the orbital angular momentum $L$ of the colliding nuclei with the friction coefficients $\gamma_R$ and $\gamma_\theta$ \cite{Fazio2004,Giardina2000,Kayumov2025}:
\begin{eqnarray}
&&\mu(R)\ddot{R}+\gamma_R(R,\alpha_1,\alpha_2)\dot{R}(t)=F(R,\alpha_1,\alpha_2), \nonumber\\
&&F(R,\alpha_1,\alpha_2)=-\frac{\partial V(R,L,\alpha_1,\alpha_2)}{\partial R}-\dot{R}^2\frac{\partial\mu(R)}{\partial R},\nonumber\\
&&\frac{dL}{dt}=-\gamma_\theta(R,\alpha_1,\alpha_2)R(t)\nonumber\\
&&\qquad\times\left(\dot{\theta}R(t)-\dot{\theta}_1R_{1\rm eff}-\dot{\theta}_2R_{2\rm eff}\right),
\label{eq:eom}
\end{eqnarray}
where $\alpha_1$ and $\alpha_2$ are the orientation angles of the axial symmetry axes of the projectile and target nuclei relative to the beam direction, $\dot\theta$ is the angular velocity of the relative motion, $\dot\theta_1$ and $\dot\theta_2$ are the angular velocities of the rotation of the fragments around their centers, and $R_{i\rm eff}$ are the effective radii of the fragments at the contact point. The reduced mass is $\mu(R)=m_NA_1A_2/(A_1+A_2)+\delta\mu(R)$, where the correction $\delta\mu$ caused by the overlap of the nuclear densities is less than 1\% of $\mu$; therefore, the term with the derivative of the reduced mass has a negligible effect on capture \cite{Fazio2004}. The radial and tangential friction coefficients $\gamma_R$ and $\gamma_\theta$ are calculated microscopically from the coupling of the relative motion with the particle-hole excitations and the nucleon exchange between the nuclei \cite{Giardina2000}. The orbital angular momentum $L$ lost by the relative motion is transferred to the rotation of the fragments, so that $L+L_1+L_2=L_0$, and $L$ does not decrease below the rolling limit $5L_0/7$. The nucleus-nucleus potential $V(R,L,\alpha_1,\alpha_2)$ is the sum of the Coulomb, nuclear, and rotational potentials, and its nuclear part is calculated by the folding procedure with the effective nucleon-nucleon forces of Migdal \cite{Migdal1967}. The target nuclei are treated as rotors with the static deformation parameters $\beta_2=0.250$, $\beta_3=0.045$, and $\beta_4=0.039$ for $^{249}$Cf and $\beta_2=0.250$, $\beta_3=0.055$, and $\beta_4=0.039$ for $^{249}$Bk. For $^{249}$Cf, and $^{249}$Bk, $\beta_2$ is the FRDM2012 value \cite{Moller2016}, and $\beta_3$ for $^{249}$Bk is the mean value of the $N=152$ isotones $^{248}$Cm and $^{250}$Cf, and  for $^{249}$Cf mean value of the $^{248}$Cf and $^{250}$Cf is used from \cite{Kibedi2002}; $\beta_4$ is the FRDM2012 value for both nuclei \cite{Moller2016}. Since $\beta_3$ is small, the difference between the orientations $\alpha_2$ and $180^\circ-\alpha_2$ is neglected. The spherical projectile $^{50}$Ti is treated through the zero-point vibrations of its first $2^+$ and $3^-$ states \cite{Raman2001,Kibedi2002}.

The partial capture cross section is calculated as
\begin{equation}
\sigma_{\rm cap}(E_{\rm c.m.},L,\alpha_i)=\pi\lambdabar^2(2L+1)P^{(L)}_{\rm cap}(E_{\rm c.m.},\alpha_i),
\label{eq:scap}
\end{equation}
where $\lambdabar$ is the de Broglie wavelength of the entrance channel. Capture occurs when two conditions are satisfied \cite{Kayumov2025}:
\begin{enumerate}
\item the collision energy is sufficient to reach the potential well of the nucleus-nucleus interaction by overcoming or tunneling through the entrance-channel barrier;
\item the kinetic energy dissipated during the approach is large enough for the system to be trapped in the potential well.
\end{enumerate}
In the present calculations, the potential well and the barrier, $V_{\rm min}$ and $V_B$, are determined from the potential $V(R,L(t),\alpha_1,\alpha_2)$ calculated with the current angular momentum and the orientations of the nuclei along the trajectory. The trajectory is followed through its turning points until the system either crosses the barrier radius $R_B$ on the way out or stays inside it. The capture probability $P^{(L)}_{\rm cap}$ is equal to 1 if the trajectory has reached the region $R<R_B$, ends inside it, and the total energy of the relative motion remaining after the dissipation, $E_{\rm tot}$, is lower than $V_B+\hbar\omega_B$, where $\hbar\omega_B$ is the curvature of the barrier; otherwise $P^{(L)}_{\rm cap}=0$. If the trajectory is reflected before the barrier, the capture probability is given by the penetrability of the parabolic barrier, $P^{(L)}_{\rm cap}=\{1+\exp[2\pi(V_B-E_{\rm tot})/\hbar\omega_B]\}^{-1}$, calculated with the energy $E_{\rm tot}$ that the system has at the barrier. Capture does not occur if the potential has no well or if its minimum lies above $E_{\rm c.m.}$.

\subsection{Rotation of the DNS}
\label{sec:rotation}
During the approach, the orbital angular momentum $L(t)$ of Eq.~(\ref{eq:eom}) enters the centrifugal part of $V$. After capture, the DNS is assumed to reach the sticking limit and to rotate as a rigid body with the total angular momentum $L_0$, which is shared between the orbital motion and the rotation of the fragments. The rotational energy of the DNS is
\begin{eqnarray}
V_{\rm rot}(L_0)&=&\frac{\hbar^2L_0(L_0+1)}{2\mathcal{J}_{\rm DNS}},\nonumber\\
\mathcal{J}_{\rm DNS}&=&\mu R^2+\mathcal{J}_1+\mathcal{J}_2,\qquad \mathcal{J}_i=\frac{M_i}{5}\left(a_i^2+c_i^2\right),
\label{eq:jdns}
\end{eqnarray}
where $a_i$ and $c_i$ are the semiaxes of the fragments. The rotational energy of the compound nucleus is $\hbar^2L_0(L_0+1)/2\mathcal{J}_{\rm CN}$ with the rigid-body moment of inertia of the spherical nucleus, $\mathcal{J}_{\rm CN}=\frac25m_Nr_0^2A_{\rm CN}^{5/3}$, $r_0=1.18$ fm. In this convention the orbital angular momentum of the DNS is $(\mu R^2/\mathcal{J}_{\rm DNS})L_0$, the potential energy surface and the quasifission barrier depend on $L_0$ but not on $E_{\rm c.m.}$, and one calculation for each $L_0$ and target orientation is used for all collision energies. Below, $L$ denotes the total angular momentum $L_0$ of the DNS. It should be noted that the potential well used in the capture test (Sec.~\ref{sec:capture}) is calculated with $L(t)$, while the quasifission barrier of the DNS is calculated in the sticking limit, i.e., it is assumed that this limit is reached after capture.

\subsection{Averaging over orientations}
\label{sec:avg}
The cross sections are averaged over the orientation angle $\alpha_2$ of the target and over the zero-point vibrations of the projectile:
\begin{equation}
\langle\sigma(E_{\rm c.m.},L)\rangle=\sum_{k,m} w^{(1)}_k\,w^{(2)}_m\,\sigma(E_{\rm c.m.},L,\theta^{(1)}_k,\alpha^{(2)}_m),
\label{eq:avg}
\end{equation}
where $w^{(2)}_m$ are the $\sin\alpha_2$ trapezoid weights on the grid $\alpha_2=0^\circ,15^\circ,\dots,90^\circ$. The surface of $^{50}$Ti is described by the quadrupole and octupole vibration amplitudes $\beta_\lambda(\theta^{(1)})=\beta^{\rm vib}_\lambda\cos2\theta^{(1)}$ ($\lambda=2,3$) with the zero-point amplitudes $\beta^{\rm vib}_2=0.166$ and $\beta^{\rm vib}_3=0.170$ \cite{Raman2001,Kibedi2002}, on the same grid of the phase $\theta^{(1)}=0^\circ,15^\circ,\dots,90^\circ$. The weights $w^{(1)}_k$ are those of the piecewise-linear quadrature in the variable $\cos2\theta^{(1)}$ with the zero-point Gaussian distribution $g\propto\exp[-2\pi Y_{20}^2(\beta_2/\beta^{\rm vib}_2)^2]$ of Eqs.~(18)--(20) of Ref.~\cite{Kayumov2022}, where $Y_{20}=\sqrt{5/4\pi}$. In total, $7\times7$ configurations are calculated for every energy.

\subsection{Capture cross sections}
\label{sec:capres}

\begin{figure*}[!tbp]
\includegraphics[width=0.9\textwidth]{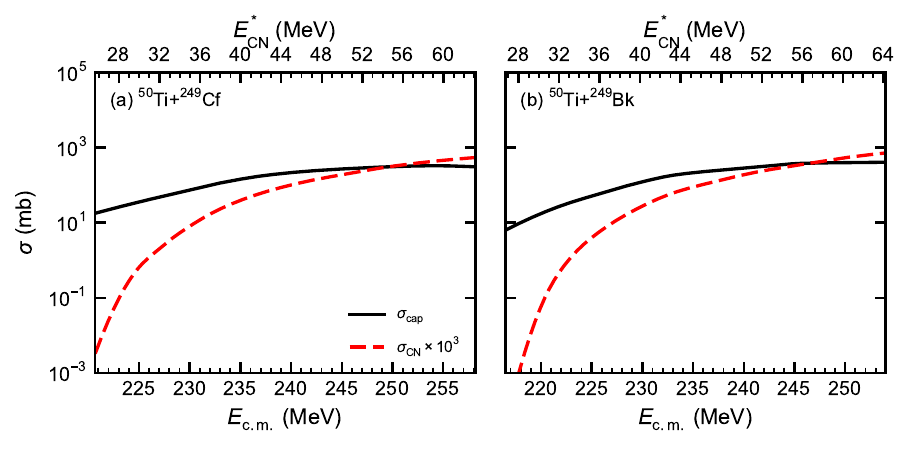}
\caption{Capture (solid curves) and fusion (dashed curves, multiplied by $10^3$) cross sections of the (a) $^{50}$Ti+$^{249}$Cf and (b) $^{50}$Ti+$^{249}$Bk reactions, averaged over the orientations of the target and the zero-point vibrations of $^{50}$Ti, as functions of $E_{\rm c.m.}$ (bottom axis) and $E^*_{\rm CN}$ (top axis). The fusion cross sections are calculated with the adopted definition (iii) of the fusion probability.}
\label{fig:cap}
\end{figure*}

The calculations have been performed for the beam energies $E_{\rm lab}=265$--310 MeV for $^{50}$Ti+$^{249}$Cf and 260--305 MeV for $^{50}$Ti+$^{249}$Bk with the step of 5 MeV. The excitation energy of the compound nucleus is $E^*_{\rm CN}=E_{\rm c.m.}+Q_{gg}$, with $Q_{gg}=-195.03$ and $-189.78$ MeV, respectively, so that $E^*_{\rm CN}=25.7$--63.1 MeV and 26.7--64.2 MeV. To locate the maxima of the 3n excitation functions, the calculations have also been performed at $E_{\rm lab}=272.5$ MeV ($E^*_{\rm CN}=31.9$ MeV) and 267.5 MeV ($E^*_{\rm CN}=33.0$ MeV), respectively. In Fig.~\ref{fig:cap} and in the figures of the fusion probability, the survival probability, and the ER cross sections below, the curves are shape-preserving cubic spline interpolations of the logarithm of the values calculated at these energies and on the $L$ grid with step $5\hbar$.

Figure~\ref{fig:cap} shows the capture cross sections averaged by Eq.~(\ref{eq:avg}) over the orientations of the target and the zero-point vibrations of $^{50}$Ti (solid curves). At the lowest energies only the collisions with the tip region of the target ($\alpha_2\le45^\circ$) lead to capture. It should be noted that the angular momentum window of capture is determined dynamically. For some orientations the trajectories with small $L$ are reflected from a shallow pocket, and for some configurations (for example, $\theta^{(1)}=90^\circ$, $\alpha_2=90^\circ$) the potential has no pocket in the sudden approximation and capture does not occur. These features are kept as calculated. For some configurations at high energies (19 of the 49 configurations of $^{50}$Ti+$^{249}$Cf at $E_{\rm lab}=290$ MeV), the capture stops at $L=55$--$100\hbar$, but the trajectory calculation gives capture again at higher $L$, with the trajectories trapped in very shallow wells ($\le0.3$ MeV) of the potential calculated along the trajectory. Since the capture window is expected to close with increasing $L$, this branch is considered as an artifact of the search for the potential well at very large $L$ and is excluded from the calculation. The ER cross sections are not affected, since the survival probability is negligible for $L>35\hbar$.

\section{COMPOUND NUCLEUS FORMATION PROBABILITY}
\label{sec:cn}

\subsection{Driving potential and fusion barriers}
\label{sec:drv}
The potential energy of the DNS with the light fragment $(Z,A)$ and the heavy fragment $(Z_{\rm CN}-Z,A_{\rm CN}-A)$ is calculated as \cite{Kayumov2025,Nasirov2024}
\begin{equation}
U(Z,A,L)=V(Z,A,L,R_m)+Q_{gg}(Z,A)-V^{\rm CN}_{\rm rot}(L),
\label{eq:pes}
\end{equation}
for every orientation $\alpha_2$ of the target, where $Q_{gg}(Z,A)=B_1+B_2-B_{\rm CN}$ is the reaction balance energy of the DNS fragments, with the binding energies taken negative, $B_i=M_i-Z_im_p-N_im_n<0$, $V(Z,A,L,R_m)$ includes the rotational energy of Eq.~(\ref{eq:jdns}), and $R_m$ is the position of the minimum of the potential well. The ground-state binding energies are used, i.e., the temperature dependence of the shell corrections of the fragments is neglected. For every charge number $Z$ of the light fragment, the driving potential $U_{\rm dr}(Z,L)$ is the value of Eq.~(\ref{eq:pes}) for the isotope with the lowest energy. The deformation parameters of a fragment are taken from the table of Ref.~\cite{Moller1995} only if the fragment is a rigid prolate rotor, i.e., if $A\ge50$, $\beta_2\ge0.15$, and the table values of $\beta_2$ for $A\pm1$ are also positive; otherwise the fragment is taken spherical. In particular, the isotopes $^{50-54}$Ti have $\beta_2=0$ in this table. As a result, the vibration phase of $^{50}$Ti does not enter the driving potential, which depends only on the orientation $\alpha_2$ of the target; the orientation of the heavy fragment is assumed to be frozen at $\alpha_2$ for all $Z$.

The intrinsic fusion barrier and the symmetry barrier of the DNS with the charge configuration $Z$ are determined by the maxima of the driving potential on the way to the compound nucleus and to the symmetric configuration, respectively \cite{Kayumov2022,NasirovKayumov2024}:
\begin{eqnarray}
B^*_{\rm fus}(Z)&=&\max_{Z'\le Z}U_{\rm dr}(Z')-U_{\rm dr}(Z),\nonumber\\
B^*_{\rm sym}(Z)&=&\max_{Z\le Z'\le Z_{\rm CN}/2}U_{\rm dr}(Z')-U_{\rm dr}(Z);
\label{eq:bfus}
\end{eqnarray}
both barriers are non-negative, and $B^*_{\rm sym}=0$ if the driving potential decreases from $Z$ toward the symmetric configurations. The position of the maximum of $U_{\rm dr}$ on the way to the compound nucleus is denoted by $Z_{\rm max}$. The quasifission barrier $B_{\rm qf}(Z,L)=V_{\rm bar}(R_{\rm qf})-V_{\rm min}(R_m)$ is the depth of the potential well of the nucleus-nucleus interaction for the configuration $Z$ and the angular momentum $L$ \cite{Nasirov2013}.

\begin{table*}[!tbp]
\caption{Driving potential $U_{\rm dr}(Z_P)$ of the DNS with the light fragment $Z_P=22$, quasifission barrier $B_{\rm qf}$, intrinsic fusion barrier $B^*_{\rm fus}$ at $L=0$, $30\hbar$, and $60\hbar$, symmetry barrier $B^*_{\rm sym}$, and the position $Z_{\rm max}$ of the maximum of the driving potential on the way to the compound nucleus for the orientation angles $\alpha_2$ of the target in the $^{50}$Ti+$^{249}$Cf reaction; $U_{\rm dr}$, $B_{\rm qf}$, $B^*_{\rm sym}$, and $Z_{\rm max}$ are given for $L=0$.}
\label{tab:barriers}
\begin{tabular}{rrrrrrrc}
\hline\hline
$\alpha_2$ & $U_{\rm dr}(Z_P)$ & $B_{\rm qf}$ & $B^*_{\rm fus}(0)$ & $B^*_{\rm fus}(30\hbar)$ & $B^*_{\rm fus}(60\hbar)$ & $B^*_{\rm sym}$ & $Z_{\rm max}$ \\
(deg) & (MeV) & (MeV) & (MeV) & (MeV) & (MeV) & (MeV) & \\
\hline
0 & 1.48 & 2.16 & 18.59 & 19.38 & 21.72 & 0.00 & 9 \\
15 & 3.69 & 2.11 & 16.36 & 17.12 & 19.36 & 0.00 & 9 \\
30 & 9.29 & 2.11 & 10.78 & 11.49 & 13.53 & 0.00 & 9 \\
45 & 15.11 & 2.35 & 6.49 & 6.64 & 8.32 & 0.49 & 17 \\
60 & 19.01 & 2.73 & 6.08 & 6.24 & 6.76 & 0.44 & 17 \\
75 & 20.64 & 2.61 & 7.07 & 7.28 & 8.24 & 0.00 & 17 \\
90 & 20.98 & 2.31 & 7.59 & 7.81 & 8.95 & 0.00 & 17 \\
\hline\hline
\end{tabular}

\end{table*}

\begin{figure}[!tbp]
\includegraphics[width=\columnwidth]{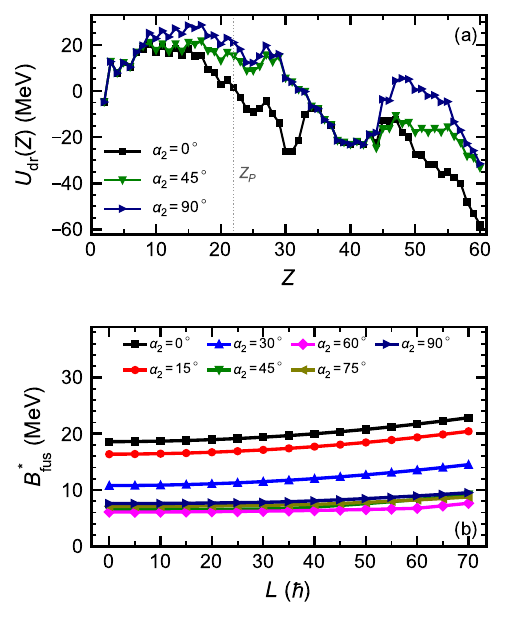}
\caption{(a) Driving potential $U_{\rm dr}(Z)$ of the DNS formed in the $^{50}$Ti+$^{249}$Cf reaction at $L=0$ for the orientation angles $\alpha_2=0^\circ$, $45^\circ$, and $90^\circ$ of $^{249}$Cf; the dotted line marks the charge number $Z_P$ of the projectile. (b) Intrinsic fusion barrier $B^*_{\rm fus}$ of the DNS with $Z=Z_P$ as a function of the angular momentum $L$ for the orientation angles $\alpha_2$ of $^{249}$Cf.}
\label{fig:bfus}
\end{figure}

Table~\ref{tab:barriers} and Fig.~\ref{fig:bfus} present the driving potential and the barriers of the DNS formed in the $^{50}$Ti+$^{249}$Cf reaction for the seven orientations of the target. Near the entrance channel the driving potential has a local minimum at $Z=24$--25, and it decreases strongly for $Z\ge30$ toward the symmetric configurations [Fig.~\ref{fig:bfus}(a)]. Its maximum on the way to the compound nucleus is located at $Z_{\rm max}=9$ for the polar orientations ($\alpha_2\le30^\circ$) and at $Z_{\rm max}=17$ for $\alpha_2\ge45^\circ$. It can be seen that the intrinsic fusion barrier depends strongly on the orientation: it is much higher for the collisions with the tip of $^{249}$Cf than for the equatorial orientations, since for the polar orientations the DNS lies low on the driving potential. The quasifission barrier of the initial configuration is 2.1--2.7 MeV for all orientations, and the symmetry barrier is zero or small. The fusion barrier increases with $L$, and the increase is largest for the polar orientations [Fig.~\ref{fig:bfus}(b)]. The driving potential of the DNS formed in the $^{50}$Ti+$^{249}$Bk reaction has a similar structure.

\subsection{Evolution of the charge distribution}
\label{sec:dzevol}
After capture, the DNS evolves by the transfer of nucleons between its fragments and can decay into two fragments (quasifission) from any configuration. The probability $D_Z(t)$ of the charge configuration $Z$ is obtained from the solution of the transport master equation \cite{Kayumov2022,Nasirov2005,Adamian1997}
\begin{eqnarray}
\frac{dD_Z(t)}{dt}&=&\sum_{Z'=Z\pm1}\left[\lambda_{Z'\to Z}D_{Z'}(t)-\lambda_{Z\to Z'}D_Z(t)\right]\nonumber\\
&&-\Lambda^{\rm qf}_ZD_Z(t),
\label{eq:master}
\end{eqnarray}
with the initial condition corresponding to the entrance channel, $D_Z(0)=\delta_{Z,Z_P}$. The equation is integrated with the time step $10^{-22}$ s. The microscopic rates $\Lambda_{i\to j}$ of the transfer of one nucleon between the configurations $i$ and $j$ of the DNS are calculated with the single-particle schemes of the fragments as in Refs.~\cite{Kayumov2022,Nasirov2005} and are symmetrized as
\begin{equation}
\lambda_{i\to j}=\sqrt{\Lambda_{i\to j}\Lambda_{j\to i}}\,\sqrt{\rho_j/\rho_i},\qquad \rho_i=\exp\left(2\sqrt{aE^*_i}\right),
\label{eq:db}
\end{equation}
with $a=A_{\rm CN}/12$ MeV$^{-1}$, so that the transition rates satisfy detailed balance, $\rho_i\lambda_{i\to j}=\rho_j\lambda_{j\to i}$, and the drift of the DNS follows the gradient of the potential energy \cite{Adamian1997}. The symmetrization keeps the geometric mean $\sqrt{\Lambda_{i\to j}\Lambda_{j\to i}}$ of the microscopic rates, i.e., their time scale, and changes only their ratio. Here $E^*_i$ is the excitation energy of the DNS in the configuration $i$ with the potential energy $U_i$,
\begin{equation}
E^*_i=E_{\rm c.m.}-V_{\rm min}(Z_P,A_P)-\left[U_i-U(Z_P,A_P)\right].
\label{eq:estar}
\end{equation}

The decay of every configuration by quasifission is described by the Kramers rate \cite{Kramers1940,Nasirov2013}
\begin{eqnarray}
\Lambda^{\rm qf}_i&=&\frac{\omega_m}{2\pi\omega_{\rm qf}}\left(\sqrt{\left(\frac{\Gamma}{2\hbar}\right)^2+\omega_{\rm qf}^2}-\frac{\Gamma}{2\hbar}\right)\nonumber\\
&&\times\exp\left(-\frac{B_{\rm qf}(i)}{T_i}\right),
\label{eq:kramers}
\end{eqnarray}
where $\omega_m$ and $\omega_{\rm qf}$ are the frequencies of the harmonic oscillator approximating the nucleus-nucleus potential $V(R)$ of the configuration $i$ at the minimum and at the quasifission barrier, obtained from the curvatures of $V(R)$ and the reduced mass at the given $L$, and $T_i=\sqrt{E^*_i/a}$ with $a=A_{\rm CN}/12$ MeV$^{-1}$. The width $\Gamma=2$ MeV characterizes the friction in the quasifission degree of freedom, $\Gamma/\hbar=\gamma/\mu$ in the notation of Ref.~\cite{Nasirov2013}. The configurations without a potential well decay by quasifission immediately. The configurations beyond the maximum of the driving potential, for which the driving potential lies at least 5 MeV below this maximum, are treated as the compound nucleus. Thus, the probability of the DNS, $P_{\rm DNS}(t)=\sum_ZD_Z(t)$, decreases with time by quasifission, $Y_{\rm qf}(t)$, and by the transitions to these configurations, $P_{\rm fus}(t)$, with $P_{\rm DNS}+Y_{\rm qf}+P_{\rm fus}=1$. Within $t_0=2\times10^{-21}$ s, the DNS spreads from the entrance configuration over the local minima of the driving potential near $Z_P$, while the probability of the configurations at and beyond the maximum of the driving potential remains small and increases slowly with time.

\subsection{Fusion probability}
\label{sec:pcn}
The fusion probability is calculated as a sum over the charge configurations of the DNS, in each of which complete fusion competes with quasifission \cite{Kayumov2022,NasirovKayumov2024}:
\begin{equation}
P_{\rm CN}(E_{\rm c.m.},L,\alpha_2)=\sum_{Z}\tilde D_Z(t_0)\,P^{(Z)}_{\rm CN}(E^*_Z,L,\alpha_2),
\label{eq:pcn}
\end{equation}
where the sum runs up to the symmetric configuration $Z_{\rm CN}/2$, and $\tilde D_Z(t_0)$ is the charge distribution of the DNS surviving at the time $t_0$. Since the DNS decays by quasifission during its evolution, the solution of Eq.~(\ref{eq:master}) is not normalized to unity, $\sum_ZD_Z(t)=P_{\rm DNS}(t)<1$, and it is divided by the probability that the DNS has not decayed by the time $t_0$:
\begin{equation}
\tilde D_Z(t_0)=\frac{D_Z(t_0)}{P_{\rm DNS}(t_0)},\qquad \sum_Z\tilde D_Z(t_0)=1.
\label{eq:norm}
\end{equation}
Thus, only the shape of the charge distribution, i.e., the population of the configurations by the surviving DNS, enters Eq.~(\ref{eq:pcn}), while the part of the DNS that has decayed before $t_0$ is not taken into account explicitly. The fusion probability of the configuration $Z$ is the branching ratio of the level densities over the intrinsic fusion barrier, the quasifission barrier, and the symmetry barrier \cite{NasirovKayumov2024}:
\begin{equation}
P^{(Z)}_{\rm CN}(E^*_Z)=\frac{\rho_{\rm fus}(E^*_Z)}{\rho_{\rm fus}(E^*_Z)+\rho_{\rm qf}(E^*_Z)+\rho_{\rm sym}(E^*_Z)},
\label{eq:branch}
\end{equation}
where $\rho_k(E^*_Z)$ is the Fermi-gas level density at the excitation energy $E^*_Z-B_k(Z)$ above the corresponding barrier, $B^*_{\rm fus}$, $B_{\rm qf}$, or $B^*_{\rm sym}$, with the level density parameter $a=A_{\rm CN}/10$ MeV$^{-1}$ \cite{Nasirov2013}, and $\rho_k=0$ if $E^*_Z\le B_k$. The configurations without a potential well ($B_{\rm qf}=0$) do not contribute to fusion. The excitation energy $E^*_Z$ is given by Eq.~(\ref{eq:estar}) with $U_i=U_{\rm dr}(Z,L)$.

In this approach, the transport calculation determines the population of the charge configurations that the DNS reaches during the charge equilibration, while the competition between complete fusion and quasifission in each configuration is taken into account statistically by Eq.~(\ref{eq:branch}). The statistical treatment is justified by the short thermalization time of the DNS, $\tau_{\rm ther}\approx2.6\times10^{-22}/T^2$ s \cite{Fazio2004}, i.e., $(1$--$7)\times10^{-22}$ s at the present excitation energies, which is shorter than $t_0$. Equation~(\ref{eq:branch}) describes the whole competition between complete fusion and quasifission in the configuration $Z$, including the quasifission that has occurred before $t_0$. If the decayed part of the DNS were also kept, i.e., if $D_Z(t_0)$ were used in Eq.~(\ref{eq:pcn}) without the normalization of Eq.~(\ref{eq:norm}), quasifission would be counted twice, once by the Kramers rate in the transport calculation and once by Eq.~(\ref{eq:branch}). With the normalization, quasifission is taken into account once, and the Kramers rate of Eq.~(\ref{eq:kramers}) only shapes the charge distribution.

The charge distribution in Eq.~(\ref{eq:pcn}) is taken at $t_0=2\times10^{-21}$ s. During the first $10^{-21}$ s, the normalized charge distribution changes rapidly as the DNS spreads from the entrance configuration; after $t_0$ it changes by less than 1\% per $10^{-22}$ s, i.e., $D_Z$ has acquired a quasi-stationary shape. This value of $t_0$ is of the same order as the interaction time $10^{-21}$ s used in Ref.~\cite{Kayumov2022}. The part of the DNS that reaches the compound nucleus configurations of the transport calculation before $t_0$, $P_{\rm fus}(t_0)$, is negligible in comparison with $P_{\rm CN}$. The partial fusion cross section is then calculated as
\begin{eqnarray}
\sigma_{\rm CN}(E_{\rm c.m.},L)&=&\sum_{k,m}w^{(1)}_kw^{(2)}_m\,\sigma_{\rm cap}(E_{\rm c.m.},L,\theta^{(1)}_k,\alpha^{(2)}_m)\nonumber\\
&&\times P_{\rm CN}(E_{\rm c.m.},L,\alpha^{(2)}_m),
\label{eq:scn}
\end{eqnarray}
where $P_{\rm CN}$ is calculated on the $L$ grid with step $5\hbar$ and interpolated log-linearly between the nodes.

\begin{figure*}[!tbp]
\includegraphics[width=0.9\textwidth]{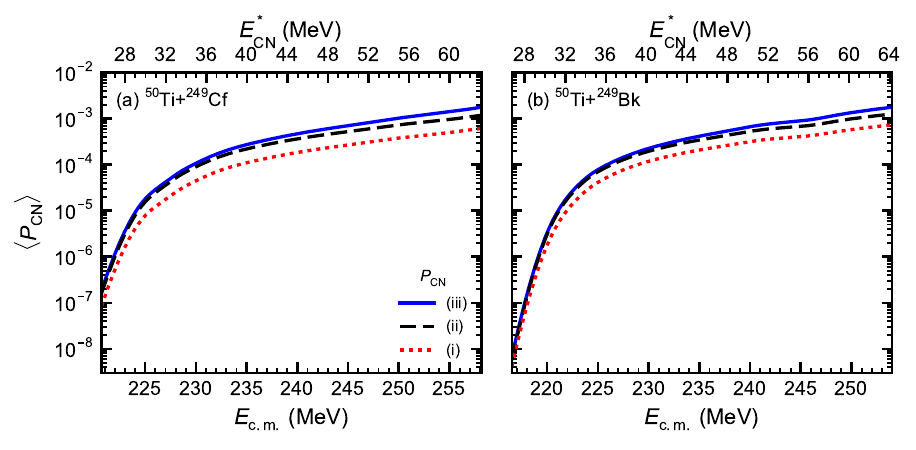}
\caption{Fusion probabilities $\langle P_{\rm CN}\rangle$ of the (a) $^{50}$Ti+$^{249}$Cf and (b) $^{50}$Ti+$^{249}$Bk reactions obtained in the three ways described in the text, averaged over the orientations of the target and the zero-point vibrations of $^{50}$Ti, as functions of $E_{\rm c.m.}$ (bottom axis) and $E^*_{\rm CN}$ (top axis).}
\label{fig:pcn}
\end{figure*}

The same transport calculation makes it possible to estimate the fusion probability in three ways, which differ in the treatment of the competition between complete fusion and quasifission. (i) In the transport calculation alone, the compound nucleus configurations absorb the part of the DNS that reaches them by diffusion over the maximum of the driving potential, while quasifission removes the rest. The probability $P_{\rm fus}(t)$ increases until the whole DNS has decayed in one of these two ways, and its limit $P_{\rm fus}(t\to\infty)$ is a purely dynamical estimate of the fusion probability. Since the diffusion over the maximum of the driving potential is slow and the transition rates do not change after $10^{-20}$ s, this limit is obtained from the DNS remaining at the end of the calculation, $t=2\times10^{-20}$ s, by solving the master equation with these rates as an absorbing Markov chain. (ii) Equation~(\ref{eq:pcn}) with the charge distribution $D_Z(t_0)$ instead of $\tilde D_Z(t_0)$: the quasifission before $t_0$ is taken from the transport calculation and the competition after $t_0$ from Eq.~(\ref{eq:branch}), i.e., the two stages are counted sequentially, and $P_{\rm CN}$ is smaller by the factor $P_{\rm DNS}(t_0)$. As discussed above, in this case the quasifission before $t_0$ is counted twice. (iii) Equation~(\ref{eq:pcn}) with the normalized charge distribution of Eq.~(\ref{eq:norm}), which is adopted in this work: the transport calculation gives only the population of the configurations, and Eq.~(\ref{eq:branch}) gives the whole competition.

\begin{table}[!tbp]
\caption{Fusion probability obtained from the same transport calculation in three ways, averaged over the orientations and the partial waves with the capture cross sections: (i) the probability $P_{\rm fus}(t\to\infty)$ absorbed by the compound nucleus configurations of the transport calculation, (ii) Eq.~(\ref{eq:pcn}) with $D_Z(t_0)$ instead of $\tilde D_Z(t_0)$ (not normalized), and (iii) Eq.~(\ref{eq:pcn}) with the normalized $\tilde D_Z(t_0)$ (adopted); $t_0=2\times10^{-21}$ s. The last column is the ratio (ii)/(iii).}
\label{tab:pcndefs}
\begin{tabular}{rcccc}
\hline\hline
$E^*_{\rm CN}$ & (i) & (ii) & (iii) & (ii)/(iii) \\
(MeV) & $P_{\rm fus}(t\to\infty)$ & not norm. & adopted & \\
\hline
\multicolumn{5}{l}{$^{50}$Ti+$^{249}$Cf} \\
25.7 & $8.4\times10^{-8}$ & $1.6\times10^{-7}$ & $1.8\times10^{-7}$ & 0.87 \\
29.8 & $7.2\times10^{-6}$ & $1.4\times10^{-5}$ & $1.7\times10^{-5}$ & 0.86 \\
34.0 & $3.5\times10^{-5}$ & $7.0\times10^{-5}$ & $8.3\times10^{-5}$ & 0.85 \\
38.2 & $8.6\times10^{-5}$ & $1.7\times10^{-4}$ & $2.1\times10^{-4}$ & 0.82 \\
46.5 & $2.1\times10^{-4}$ & $4.1\times10^{-4}$ & $5.4\times10^{-4}$ & 0.77 \\
54.8 & $3.7\times10^{-4}$ & $7.3\times10^{-4}$ & $1.0\times10^{-3}$ & 0.72 \\
\multicolumn{5}{l}{$^{50}$Ti+$^{249}$Bk} \\
30.9 & $3.7\times10^{-6}$ & $6.5\times10^{-6}$ & $7.0\times10^{-6}$ & 0.92 \\
35.1 & $3.9\times10^{-5}$ & $6.7\times10^{-5}$ & $7.4\times10^{-5}$ & 0.90 \\
39.2 & $1.0\times10^{-4}$ & $1.7\times10^{-4}$ & $1.9\times10^{-4}$ & 0.87 \\
43.4 & $1.7\times10^{-4}$ & $2.9\times10^{-4}$ & $3.4\times10^{-4}$ & 0.84 \\
51.7 & $3.5\times10^{-4}$ & $5.9\times10^{-4}$ & $7.5\times10^{-4}$ & 0.78 \\
60.1 & $5.7\times10^{-4}$ & $9.8\times10^{-4}$ & $1.3\times10^{-3}$ & 0.73 \\
\hline\hline
\end{tabular}

\end{table}

The fusion cross sections are shown in Fig.~\ref{fig:cap}, and the three estimates of the fusion probability in Fig.~\ref{fig:pcn} and in Table~\ref{tab:pcndefs}. The adopted $\langle P_{\rm CN}\rangle$ exceeds $P_{\rm fus}(t\to\infty)$ by a factor of 2.2--2.9 for $^{50}$Ti+$^{249}$Cf and 1.6--2.4 for $^{50}$Ti+$^{249}$Bk at all energies, and the value without the normalization lies between the two. Thus, the two parts of the model, the transport over the driving potential and the statistical branching of Eq.~(\ref{eq:branch}), give fusion probabilities of the same order, and their difference characterizes the uncertainty of the definition of $P_{\rm CN}$. At the lowest energies only the collisions with the tip region of the target lead to capture, and the polar orientations among them have the highest fusion barriers. As a result, $\langle P_{\rm CN}\rangle$ increases steeply at the lowest energies and then slowly, reaching about $2\times10^{-3}$ at the highest energies for both reactions. At the same excitation energy above 35 MeV, the fusion probabilities of the two reactions are close to each other.

The fusion probability $P_{\rm CN}=1.12\times10^{-3}$ obtained for the $^{50}$Ti+$^{249}$Cf reaction in the DNS calculation of Ref.~\cite{Nasirov2011} is reached in the present calculation only at $E^*_{\rm CN}\approx56$ MeV. The semiempirical estimate of Ref.~\cite{Kozulin2016}, which is based on the fusion probabilities deduced from the experimental data, gives $P_{\rm CN}\approx4\times10^{-3}$ for this reaction \cite{Khuyagbaatar2020}, i.e., larger than all values in Table~\ref{tab:pcndefs}.

\subsection{Sensitivity of the fusion probability to $t_0$}
\label{sec:sens}

The dependence of the fusion probability on the time $t_0$ at which the charge distribution is taken has been studied for all collision energies. For $10^{-22}\le t_0\le4\times10^{-21}$ s, the orientation-averaged $P_{\rm CN}$ differs from its value at the adopted $t_0$ by less than 16\% for $^{50}$Ti+$^{249}$Cf and by less than 20\% for $^{50}$Ti+$^{249}$Bk (33\% at the lowest energy, $E^*_{\rm CN}=26.7$ MeV), although a large part of the DNS decays during this time. This weak dependence has two reasons: the normalization removes the decayed part of the DNS, and the configurations populated at these times have similar values of $P^{(Z)}_{\rm CN}$.

\section{SURVIVAL PROBABILITY}
\label{sec:survival}

The deexcitation of the compound nucleus is described by the statistical model, in which the emission of neutrons, protons, and $\gamma$ rays competes with fission at every step of the evaporation cascade. Since the fission barrier of the superheavy nuclei depends on the angular momentum, the population of the excited nuclei is followed both in the excitation energy $E$ and in the spin $J$, and the particle emission is treated in the HF formalism \cite{HauserFeshbach1952}.

\subsection{Evaporation cascade}
\label{sec:cascade}
Let $\mathcal{P}_{\rm C}(E,J)$ be the population of the states $(E,J)$ of the nucleus C$\,=(Z,N)$. The compound nucleus is populated at the excitation energy $E^*_{\rm CN}=E_{\rm c.m.}+Q_{gg}$ with the spin $J=L$, $\mathcal{P}_{\rm CN}(E,J)=\delta(E-E^*_{\rm CN})\,\delta_{J,L}$; the ground-state spins of the targets and the parity of the states are not taken into account. The population of the state $(E',I)$ of the nucleus C$'$ is
\begin{eqnarray}
\mathcal{P}_{{\rm C}'}(E',I)&=&\sum_{\nu}\sum_{E,J}\mathcal{P}_{\rm C'+\nu}(E,J)\,\frac{\Gamma_\nu(E,J\to E',I)}{\Gamma_{\rm tot}(E,J)}\nonumber\\
\rlap{$\displaystyle\qquad+\sum_{E>E',J}\mathcal{P}_{{\rm C}'}(E,J)\,\frac{\Gamma_\gamma(E,J\to E',I)}{\Gamma_{\rm tot}(E,J)},$}\phantom{\mathcal{P}_{{\rm C}'}(E',I)}&&
\label{eq:cascade}
\end{eqnarray}
where C$'+\nu$ is the parent nucleus that forms C$'$ by the emission of the particle $\nu$, and the second term describes the $\gamma$ transitions within the same nucleus. The populations are calculated in the order of decreasing excitation energy, so that every state is fed before it decays. The total width is
\begin{equation}
\Gamma_{\rm tot}(E,J)=\sum_\nu\Gamma_\nu(E,J)+\Gamma_\gamma(E,J)+\Gamma_f(E,J),
\label{eq:gtot}
\end{equation}
where $\Gamma_\nu(E,J\to E',I)$ is the partial width for the transition to the state $(E',I)$ of the daughter nucleus, $\nu=n,p$, and the fission width removes the population from the cascade. The emission of $\alpha$ particles is neglected. The survival probability $W^{xn}_{\rm sur}(E^*_{\rm CN},L)$ is the part of the initial population that is accumulated in the nucleus $A_{\rm CN}-x$ below the thresholds of particle emission and fission, i.e., that reaches the ground state of this nucleus by $\gamma$ emission. The channels with the emission of protons, $xn\,yp$, are obtained in the same way. The survival probabilities have been calculated for $L=0$--$60\hbar$ with step $5\hbar$ at the excitation energies of Sec.~\ref{sec:capres}, with the excitation-energy bins of 1 MeV, the kinetic-energy step of 0.4 MeV in the integrals of the widths, and up to six emitted neutrons and one proton. The widths of a bin are calculated at its excitation energy; a decay channel is open if the excitation energy left above the pairing shift, $E-\varepsilon-\Delta-S_\nu$ for particle emission or $E-\varepsilon-\Delta-B_f$ for fission, exceeds the kinetic-energy step, and the daughter state is assigned to the nearest energy bin.

\subsection{Particle and $\gamma$ emission}
\label{sec:hfwidths}
The width for the emission of the particle $\nu$ with the spin $s_\nu$ and the separation energy $S_\nu$ from the state $(E,J)$ of the nucleus C is
\begin{eqnarray}
\Gamma_\nu(E,J)&=&\frac{1}{2\pi\rho_{\rm C}(E,J)}\sum_{I}\sum_{\ell,j}\int_0^{E-S_\nu}T^{\nu}_{\ell j}(\varepsilon)\nonumber\\
&&\times\rho_{\rm R}(E-S_\nu-\varepsilon,I)\,d\varepsilon,
\label{eq:hf}
\end{eqnarray}
where $\varepsilon$ is the kinetic energy of the emitted particle, $\ell$ and $j$ are its orbital and total angular momenta, $|\ell-s_\nu|\le j\le\ell+s_\nu$, and the spin $I$ of the residual nucleus R satisfies $|J-j|\le I\le J+j$. Here $\rho_{\rm C}$ and $\rho_{\rm R}$ are the level densities of the emitting and residual nuclei. The transmission coefficients $T^\nu_{\ell j}$ are obtained by solving the Schr\"odinger equation with the global optical potential of Varner \textit{et al.} \cite{Varner1991} for neutrons and protons; for protons the Coulomb barrier is contained in $T^p_{\ell j}$. The separation energies are calculated with the FRDM2012 masses \cite{Moller2016}.

The $\gamma$-emission width is calculated as
\begin{eqnarray}
\Gamma_\gamma(E,J)&=&\frac{1}{2\pi\rho_{\rm C}(E,J)}\sum_{X\lambda}\sum_I\int_0^{E}2\pi\varepsilon_\gamma^{2\lambda+1}f_{X\lambda}(\varepsilon_\gamma)\nonumber\\
&&\times\rho_{\rm C}(E-\varepsilon_\gamma,I)\,d\varepsilon_\gamma,
\label{eq:gamma}
\end{eqnarray}
where the sum runs over the electric and magnetic transitions of multipolarities $\lambda=1$ and 2 with $|J-\lambda|\le I\le J+\lambda$. The E1 strength function $f_{E1}$ is taken in the enhanced generalized Lorentzian form \cite{KopeckyUhl1990} with two giant dipole resonance components split by the ground-state deformation; the centroid energy is $E_0=27.47A^{-1/3}+22.06A^{-1/6}$ MeV, the width $\Gamma_0=0.0269E_0^{1.91}$ MeV, and the peak cross section is given by the Thomas-Reiche-Kuhn sum rule with the enhancement 1.22 \cite{Capote2009}. The M1 and E2 strength functions are standard Lorentzians with $E_{M1}=41A^{-1/3}$ MeV, $\Gamma_{M1}=4$ MeV, $\sigma_{M1}=1$ mb, and $E_{E2}=63A^{-1/3}$ MeV, $\Gamma_{E2}=6.11-0.012A$ MeV, $\sigma_{E2}=1.5\times10^{-4}Z^2E_{E2}^2A^{-1/3}/\Gamma_{E2}$ mb, and $f_{M2}=0.307A^{-2/3}f_{E2}$ \cite{Capote2009}.

\subsection{Fission width}
\label{sec:fiswidth}
\begin{table*}[!tbp]
\caption{Fission-barrier inputs for the isotopes of elements 120 and 119 formed along the 2n--5n chains (in MeV): ground-state shell correction $\delta W_{\rm gs}$ and neutron separation energy $S_n$ of FRDM2012 \cite{Moller2016}; barrier $B_f^{(0)}$, saddle-point deformations $\beta_{20}^{\rm sd}$ and $\beta_{22}^{\rm sd}$ of the WS+YPE calculation \cite{Jachimowicz2021}, and the saddle-point shell correction of Eq.~(\ref{eq:dwsd}); and the alternative inputs of the sensitivity study: the barrier of Ref.~\cite{Moller2015} and the ground-state shell corrections of the KTUY \cite{Koura2005} and WS4 \cite{Wang2014} mass tables.}
\label{tab:bfinput}
\begin{tabular}{lrrrrrrrrr}
\hline\hline
 & \multicolumn{2}{c}{FRDM2012} & \multicolumn{4}{c}{WS+YPE} & Ref.~\cite{Moller2015} & KTUY & WS4 \\
Nucleus & $\delta W_{\rm gs}$ & $S_n$ & $B_f$ & $\beta_{20}^{\rm sd}$ & $\beta_{22}^{\rm sd}$ & $\delta W_{\rm sd}$ & $B_f$ & $\delta W_{\rm gs}$ & $\delta W_{\rm gs}$ \\
\hline
$^{299}$120 & -7.70 & 6.17 & 5.57 & 0.35 & 0.27 & -2.13 & 7.48 & -9.82 & -5.48 \\
$^{298}$120 & -7.52 & 7.99 & 5.56 & 0.35 & 0.27 & -1.96 & 7.33 & -9.75 & -5.89 \\
$^{297}$120 & -8.37 & 6.47 & 6.02 & 0.34 & 0.27 & -2.35 & 7.54 & -9.24 & -5.79 \\
$^{296}$120 & -8.24 & 7.88 & 5.79 & 0.35 & 0.27 & -2.45 & 7.69 & -8.79 & -6.23 \\
$^{295}$120 & -8.57 & 6.88 & 6.28 & 0.34 & 0.28 & -2.29 & 7.71 & -8.51 & -6.16 \\
$^{299}$119 & -8.17 & 7.09 & 5.72 & 0.36 & 0.25 & -2.45 & 7.72 & -9.66 & -5.45 \\
$^{298}$119 & -8.49 & 6.14 & 6.20 & 0.35 & 0.26 & -2.29 & 8.10 & -9.53 & -5.46 \\
$^{297}$119 & -8.32 & 7.71 & 6.20 & 0.35 & 0.26 & -2.12 & 7.94 & -9.41 & -5.93 \\
$^{296}$119 & -8.80 & 6.48 & 6.71 & 0.36 & 0.26 & -2.09 & 8.07 & -9.02 & -5.79 \\
$^{295}$119 & -8.96 & 7.57 & 6.32 & 0.36 & 0.28 & -2.64 & 8.06 & -8.70 & -6.12 \\
\hline\hline
\end{tabular}

\end{table*}

The fission width is calculated with the Bohr-Wheeler formula \cite{BohrWheeler1939} and is multiplied by the Strutinsky factor \cite{Strutinsky1973} and by the Kramers factor for the nuclear viscosity \cite{Kramers1940}:
\begin{eqnarray}
\Gamma_f(E,J)&=&\frac{K}{2\pi\rho_{\rm g}(E,J)}\int_0^{E-B_f}\rho_{\rm s}(E-B_f-\varepsilon,J)\,d\varepsilon,\nonumber\\
K&=&\frac{\hbar\omega_{\rm gs}}{T}\left(\sqrt{1+\eta^2}-\eta\right),
\label{eq:fis}
\end{eqnarray}
where $\rho_{\rm g}$ and $\rho_{\rm s}$ are the level densities at the ground state and at the saddle point, $\varepsilon$ is the kinetic energy in the fission degree of freedom above the barrier, $\hbar\omega_{\rm gs}=\hbar\omega_{\rm sd}=1$ MeV are the curvatures of the potential at the ground state and at the saddle, and $\eta=\beta_{\rm fr}/2\omega_{\rm sd}$ with the reduced friction coefficient $\beta_{\rm fr}=5\times10^{21}$ s$^{-1}$; these values are typical of the dynamical descriptions of fission \cite{FrobrichGontchar1998}. The first factor in $K$, $\hbar\omega_{\rm gs}/T$, is the Strutinsky correction for the collective vibrations in the ground-state well, and the second one is the Kramers reduction of the stationary fission flux. Here $T=\sqrt{U/a(U)}$ is the temperature of the ground state, with $U=E-\Delta$ and the level density parameter $a(U)$ of Eq.~(\ref{eq:ignatyuk}). The transient time of fission is neglected.

The fission barriers at zero excitation energy, $B_f^{(0)}$, are taken from the macroscopic-microscopic calculation of Jachimowicz, Kowal, and Skalski \cite{Jachimowicz2021}, which uses the Woods-Saxon single-particle potential and the Yukawa-plus-exponential macroscopic energy (WS+YPE) and searches for the saddle points in the five-dimensional deformation space including the nonaxial deformation $\beta_{22}$. For the isotopes of elements 119 and 120 formed in the present reactions, and for the oganesson isotopes formed after the emission of a proton from $^{299}$119, all saddle points are triaxial with $\beta_{22}=0.25$--0.28. The barriers, the saddle-point deformations, and the shell corrections along the 2n--5n chains are given in Table~\ref{tab:bfinput}. The macroscopic (Thomas-Fermi) barrier of these nuclei \cite{MyersSwiatecki1999}, $B_{\rm TF}$, vanishes. The barrier depends on the angular momentum through its shell part, $B_f^{(0)}-B_{\rm TF}$, which is scaled as a whole by the factor $q(J)$,
\begin{eqnarray}
B_f(J)&=&B_{\rm TF}+q(J)\left(B_f^{(0)}-B_{\rm TF}\right),\nonumber\\
q(J)&=&\left[1+\exp\left(\frac{J-J_{1/2}}{\Delta J}\right)\right]^{-1},
\label{eq:bfJ}
\end{eqnarray}
with $J_{1/2}=20\hbar$ and $\Delta J=3\hbar$ \cite{Fazio2004}. For $Z\approx120$ the macroscopic barrier is zero at all $J$, and the factor $q(J)$ represents empirically the loss of the shell stabilization with increasing spin \cite{Fazio2004}. The barrier is not damped explicitly with the excitation energy. Instead, the shell correction is damped through the level density parameters at the ground state and at the saddle point [Eq.~(\ref{eq:ignatyuk})], where the shell correction at the saddle point is defined by the barrier and the ground-state shell correction $\delta W_{\rm gs}$,
\begin{equation}
\delta W_{\rm sd}=B_f^{(0)}-B_{\rm TF}+\delta W_{\rm gs}.
\label{eq:dwsd}
\end{equation}
With these definitions, the shell contributions to the barrier and to the two level densities combine into the effective barrier
\begin{equation}
B_f^{\rm eff}(U,J)\simeq B_{\rm TF}+q(J)\left(|\delta W_{\rm gs}|-|\delta W_{\rm sd}|\right)e^{-U/E_d},
\label{eq:beff}
\end{equation}
in which the excitation energy and the angular momentum act on the same quantity; this barrier decreases to the macroscopic one at high excitation energy or spin and never becomes smaller than it. In this form the shell correction is damped once, as in the heavy-ion option of the EMPIRE code, which does not damp the barrier explicitly below the temperature of 1.65 MeV and uses an angular momentum factor of the same form as $q(J)$ \cite{EMPIRE32}. The ground-state shell corrections, deformations, and neutron separation energies are taken from FRDM2012 \cite{Moller2016}, so that the fission and neutron-emission thresholds are consistent.

The saddle-point level density is multiplied by a global normalization factor $N_f$, as in the EMPIRE code \cite{EMPIRE32}. This factor accounts for the uncertainties of the Kramers factor and of the collective enhancement and the symmetry factor of the triaxial saddle point in the level density, and it is the only parameter of the survival calculation adjusted to ER data. The value $N_f=0.19$ has been fixed by the measured 3n and 4n cross sections of the $^{48}$Ca+$^{244}$Pu reaction \cite{Oganessian2004,Gates2011,SamarkRoth2023,Gates2026}, with the fusion cross sections of this reaction calculated in the DNS model \cite{KayumovAmirov}, under the condition that the geometric mean of the ratios of the calculated and measured cross sections is equal to unity; with this value, the calculated ratio of the 4n and 3n cross sections, 2.5, is also close to the measured one.

\subsection{Level density}
\label{sec:levdens}
The level density of the nucleus with the excitation energy $E$ and spin $J$ is
\begin{equation}
\rho(E,J)=K_{\rm coll}\,\frac{2J+1}{2\sqrt{2\pi}\,\sigma^3}\exp\left[-\frac{(J+1/2)^2}{2\sigma^2}\right]\rho(U),
\label{eq:rhoj}
\end{equation}
where $\sigma^2=\mathcal{J}T/\hbar^2$ is the spin cutoff parameter, $U=E-\Delta$ is the excitation energy corrected for the pairing shift $\Delta$ ($\Delta=24/\sqrt{A}$ MeV for even-even, $12/\sqrt{A}$ MeV for odd-$A$, and 0 for odd-odd nuclei), $T=\sqrt{U/a}$, and $\mathcal{J}$ is the rigid-body moment of inertia of the ground-state or saddle-point shape, used without a reduction factor. The rotational energy is not subtracted from $U$; the spin is included through the factor in Eq.~(\ref{eq:rhoj}). Therefore, the compound nucleus enters the cascade with the total excitation energy $E^*_{\rm CN}$ for every partial wave. The intrinsic level density $\rho(U)$ is calculated with the formula of Grossjean and Feldmeier \cite{GrossjeanFeldmeier1985}, which coincides with the Fermi-gas expression $\rho(U)=\sqrt{\pi}\exp(2\sqrt{aU})/(12a^{1/4}U^{5/4})$ at high excitation energy and remains finite at $U\to0$. The asymptotic level density parameter is \cite{NerloPomorska2006}
\begin{eqnarray}
\tilde{a}&=&0.092A+0.036A^{2/3}B_s+0.275A^{1/3}B_k\nonumber\\
&&-0.00146\frac{Z^2}{A^{1/3}}B_C\ \ ({\rm MeV}^{-1}),
\label{eq:atilde}
\end{eqnarray}
where $B_s$, $B_k$, and $B_C$ are the surface, curvature, and Coulomb deformation functions of the quadrupole deformation, calculated with the ground-state deformation of FRDM2012 \cite{Moller2016} and with the saddle-point deformation $\beta_{20}^{\rm sd}$ of Ref.~\cite{Jachimowicz2021}, so that the ratio $a_f/a_n$ follows from the shapes. The shell correction is included at both shapes with the damping of Ignatyuk \textit{et al.} \cite{Ignatyuk1975} and with the same angular momentum factor as in the barrier:
\begin{equation}
a(U,J)=\tilde{a}\left[1+q(J)\,\delta W\,\frac{1-e^{-U/E_d}}{U}\right],
\label{eq:ignatyuk}
\end{equation}
with $\delta W=\delta W_{\rm gs}$ at the ground state and $\delta W=\delta W_{\rm sd}$ at the saddle point, and with the damping energy $E_d=19$ MeV. The collective enhancement of the level density is included in the form in which the vibrational enhancement is present at every shape and the rotational enhancement is added for deformed shapes \cite{EMPIRE32,Capote2009},
\begin{eqnarray}
K_{\rm coll}&=&K_{\rm vib}\left[1-\varphi(\beta_2)+\varphi(\beta_2)\,K_{\rm rot}\right],\nonumber\\
\varphi(\beta_2)&=&\left[1+\exp\left(\frac{\beta_0-|\beta_2|}{\Delta\beta}\right)\right]^{-1},
\label{eq:kcoll}
\end{eqnarray}
where $K_{\rm rot}=1+(\mathcal{J}_\perp T/\hbar^2-1)\,g(U)$, $K_{\rm vib}=1+[\exp(0.0555A^{2/3}T^{4/3})-1]\,g(U)$ \cite{Capote2009}, $g(U)=\{1+\exp[(U-U_{\rm cr})/d_{\rm cr}]\}^{-1}$ fades the enhancement out at high energy, $\beta_0=0.15$, $\Delta\beta=0.04$ \cite{Zagrebaev2001}, $U_{\rm cr}=40$ MeV, and $d_{\rm cr}=10$ MeV. For the triaxial saddle points the level density is multiplied by the symmetry factor \cite{Capote2009,EMPIRE32}
\begin{equation}
f_{\rm sym}=1+\left(\sqrt{\frac{\pi}{2}\,\frac{\mathcal{J}_\parallel T}{\hbar^2}}-1\right)g(U),
\label{eq:fsym}
\end{equation}
where $\mathcal{J}_\parallel$ is the moment of inertia around the symmetry axis, $\mathcal{J}_\parallel=\mathcal{J}_0[1-\alpha^{\rm sd}_2+0.429(\alpha^{\rm sd}_2)^2+0.268(\alpha^{\rm sd}_2)^3]$ with the rigid-sphere value $\mathcal{J}_0$ and $\alpha^{\rm sd}_2=\beta_{20}^{\rm sd}\sqrt{5/4\pi}$ \cite{Vigdor1982}.

It should be noted that three level-density prescriptions are used in the different stages of the calculation. In the DNS stage, the simple Fermi-gas forms with $a=A_{\rm CN}/12$ MeV$^{-1}$ in the transport [Eqs.~(\ref{eq:kramers}) and (\ref{eq:db})] and $a=A_{\rm CN}/10$ MeV$^{-1}$ in Eq.~(\ref{eq:branch}) are kept as in the transport and branching codes of Refs.~\cite{Kayumov2022,Nasirov2013}; only the ratios of the level densities of similar configurations enter there. In the deexcitation of the compound nucleus, the absolute level densities of nuclei with different shell corrections and deformations at low excitation energies are needed, and the more detailed prescription of Eqs.~(\ref{eq:rhoj})--(\ref{eq:fsym}) is used.

\subsection{Survival probability of the compound nuclei $^{299}$120 and $^{299}$119}
\label{sec:wsurres}

\begin{figure*}[!tbp]
\includegraphics[width=0.9\textwidth]{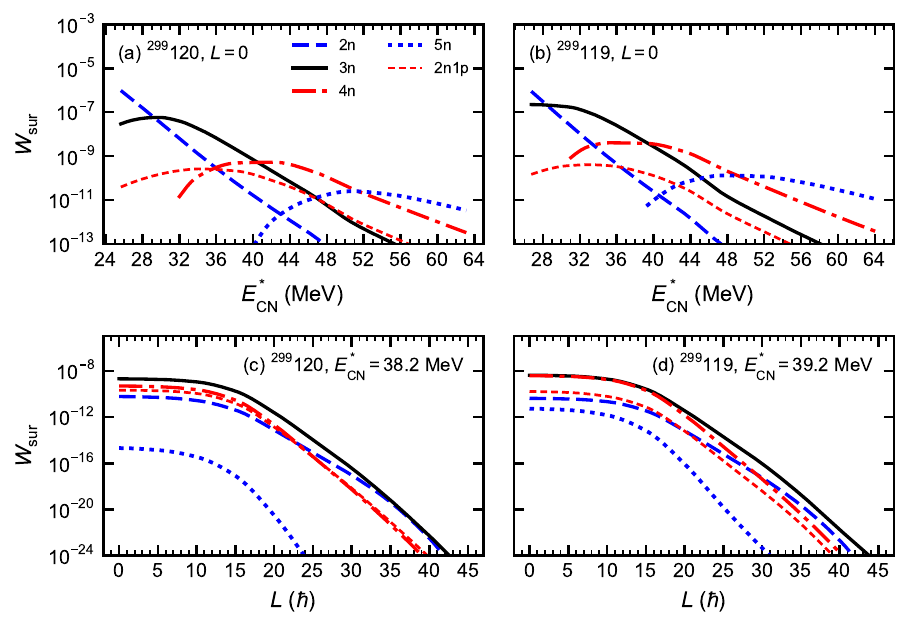}
\caption{Survival probabilities of the compound nuclei (a), (c) $^{299}$120 and (b), (d) $^{299}$119 in the 2n--5n and 2n1p channels: (a), (b) as functions of the excitation energy for $L=0$; (c), (d) as functions of $L$ at $E^*_{\rm CN}=38.2$ and 39.2 MeV, respectively.}
\label{fig:wsur}
\end{figure*}

Figures~\ref{fig:wsur}(a) and \ref{fig:wsur}(b) present the survival probabilities for $L=0$ as functions of the excitation energy. The 2n and 3n channels have the largest survival probabilities at the lowest excitation energies, while the maxima of the 4n and 5n channels are reached at higher energies. Since the barrier is damped only through the level densities, with the damping energy $E_d=19$ MeV, the survival probability decreases with the excitation energy more slowly than with the explicit thermal damping of the barrier used in our previous works \cite{Kayumov2022,Kayumov2025}, and the 4n and 5n channels are relatively stronger. The survival probabilities of $^{299}$119 in the 3n channel are larger than those of $^{299}$120 by a factor of about 4 at the same excitation energy, since the fission barriers of the isotopes of element 119 are higher with respect to their neutron separation energies (Table~\ref{tab:bfinput}). Figures~\ref{fig:wsur}(c) and \ref{fig:wsur}(d) show that the survival probability decreases with $L$ by several orders of magnitude between $L=15\hbar$ and $25\hbar$, because of the factor $q(J)$ in Eqs.~(\ref{eq:bfJ}) and (\ref{eq:ignatyuk}).

\subsection{Spin-independent cascade}
\label{sec:popcasc}
As an alternative to the cascade on the energy and spin bins described above, the survival probability has also been calculated by the method of the coupled evaporation-fission cascade \cite{KayumovAmirov}, referred to below as the spin-independent cascade. In this method the excited nuclei are characterized by their excitation energy only, and the whole energy distribution of the nuclei formed after each evaporation step is propagated, without replacing it by its mean value. Let $n_x(E)$ be the density of the nuclei that have emitted $x$ neutrons and have the excitation energy $E$, with $n_0(E)=\delta(E-E^*_{\rm CN})$. Each nucleus has three exclusive outcomes,
\begin{eqnarray}
p_{\rm res}(E)&=&\Theta(S_n-E)\,\Theta(B_f-E),\nonumber\\
p_n(E)&=&\Theta(E-S_n)\,\frac{\Gamma_n(E)}{\Gamma_n(E)+\Gamma_f(E)},\nonumber\\
p_f(E)&=&1-p_{\rm res}(E)-p_n(E),
\label{eq:pout}
\end{eqnarray}
i.e., it remains as an evaporation residue, emits a further neutron, or fissions; the second step function in $p_{\rm res}$ accounts for the nuclei that cannot emit a neutron but lie above their fission barrier. An emitted neutron takes the kinetic energy $\varepsilon$ from the spectrum
\begin{equation}
w(\varepsilon;E)=\frac{\varepsilon\exp(-\varepsilon/T_{\rm R})}{\int_0^{E-S_n}\varepsilon'\exp(-\varepsilon'/T_{\rm R})\,d\varepsilon'},
\label{eq:wspec}
\end{equation}
where $T_{\rm R}=\sqrt{U_n/a_n}$ is the temperature of the residual nucleus defined below,
so that the population after the next emission is
\begin{equation}
n_{x+1}(E')=\int dE\,n_x(E)\,p_n(E)\,w(E-S_n-E';E),
\label{eq:popcasc}
\end{equation}
and the survival probability of the $x$n channel is the part of the population that stops after exactly $x$ neutrons,
\begin{equation}
W^{xn}_{\rm sur}(E^*_{\rm CN},L)=\int n_x(E)\,p_{\rm res}(E)\,dE.
\label{eq:wpop}
\end{equation}
The ratio of the neutron and fission widths is calculated with the Vandenbosch-Huizenga expression \cite{Vandenbosch1973}, which follows from the Weisskopf neutron width \cite{Weisskopf1937} with a constant inverse cross section and the Bohr-Wheeler fission width \cite{BohrWheeler1939} with the Fermi-gas level density $\rho\propto\exp(2\sqrt{aU})$:
\begin{eqnarray}
\frac{\Gamma_n}{\Gamma_f}&=&\frac{4A_{\rm R}^{2/3}a_fU_n}{K_0a_n\left(2\sqrt{a_fU_f}-1\right)}\,\frac{1}{K}\nonumber\\
&&\times\exp\left(2\sqrt{a_nU_n}-2\sqrt{a_fU_f}\right).
\label{eq:vh}
\end{eqnarray}
Here $U_n=E-S_n-\Delta_{\rm R}$ is the excitation energy of the residual nucleus R with the mass number $A_{\rm R}$ above its pairing shift, $U_f=E-B_f-\Delta$ is the excitation energy of the emitting nucleus above the fission barrier and its pairing shift, $a_n$ is the level density parameter of Eq.~(\ref{eq:atilde}) of the residual nucleus with the Ignatyuk damping of the ground-state shell correction, $a_f=\tilde a$, and $K$ is the Kramers-Strutinsky factor of Eq.~(\ref{eq:fis}) with the temperature of the emitting nucleus; the masses, shell corrections, and pairing shifts are the same as in the HF calculation. The constant $K_0=\hbar^2/2\mu_nr_0^2$, where $\mu_n$ is the reduced mass of the neutron and the residual nucleus, corresponds to the inverse cross section $\sigma_{\rm inv}=\pi r_0^2A_{\rm R}^{2/3}$; the value $K_0=6.86$ MeV is used, which gives the same inverse cross section as the form of Dostrovsky \textit{et al.} \cite{Dostrovsky1959} with $R=1.7A^{1/3}$ fm in its high-energy limit. As in Refs.~\cite{Kayumov2022,Kayumov2025,KayumovAmirov}, the fission barrier is the macroscopic barrier corrected by the ground-state shell correction, which is damped with the temperature and the angular momentum:
\begin{eqnarray}
B_f(E^*,L)&=&B_{\rm mac}-h(T)\,q(L)\,\delta W_{\rm gs},\nonumber\\
h(T)&=&\left[1+\exp\left(\frac{T-T_0}{d}\right)\right]^{-1},
\label{eq:bfT}
\end{eqnarray}
with $T_0=1.16$ MeV, $d=0.3$ MeV \cite{Mandaglio2012,Nasirov2013}, the factor $q(L)$ of Eq.~(\ref{eq:bfJ}), and the temperature determined by the excitation energy of the nucleus at each step. The angular momentum of the compound nucleus is assumed to be conserved along the cascade and enters only through the factor $q(L)$, while the rotational energy is not subtracted from the excitation energy. A decay channel is open if $U_n$ or $U_f$ exceeds 0.4 MeV. The distributions are calculated on the grid of excitation energies with the step of 0.1 MeV up to 80 MeV. Since Eq.~(\ref{eq:popcasc}) is linear in the population, the propagation is performed for all initial energies of the grid at once, so that one pass gives $W^{xn}_{\rm sur}(E^*_{\rm CN},L)$ for every excitation energy. This form of the cascade has been tested for the $^{48}$Ca+$^{244}$Pu reaction in Ref.~\cite{KayumovAmirov}, where it agreed with the Monte Carlo solution of the same decay model.

The spin-independent cascade differs from the HF calculation of Secs.~\ref{sec:cascade}--\ref{sec:levdens} in the treatment of the evaporation cascade and of the decay widths: the neutron width is calculated with a constant inverse cross section instead of the transmission coefficients of the optical potential, the level densities are taken with the exponential factor only, without the preexponential factor, the spin dependence, and the collective enhancement, the angular momentum carried away by the neutrons is neglected, the proton and $\gamma$ emission are not included, and the thresholds of Eq.~(\ref{eq:pout}) are sharp. It also differs in the fission barrier: the shell correction at the saddle point is neglected, and the barrier is damped explicitly by Eq.~(\ref{eq:bfT}). Therefore, the comparison of the two methods shows the combined effect of the decay model and of the treatment of the shell correction in the fission barrier.

\section{EVAPORATION RESIDUE CROSS SECTIONS}
\label{sec:er}

Since Eq.~(\ref{eq:cascade}) is linear in the initial population, the ER cross section is calculated as
\begin{equation}
\sigma^{xn}_{\rm ER}(E_{\rm c.m.})=\sum_L\sigma_{\rm CN}(E_{\rm c.m.},L)\,W^{xn}_{\rm sur}(E^*_{\rm CN},L).
\label{eq:ser}
\end{equation}
The ER cross sections of the spin-independent cascade are obtained from Eq.~(\ref{eq:ser}) with the same partial fusion cross sections as in the HF calculation.

\subsection{Excitation functions}
\label{sec:erres}

\begin{figure*}[!tbp]
\includegraphics[width=0.9\textwidth]{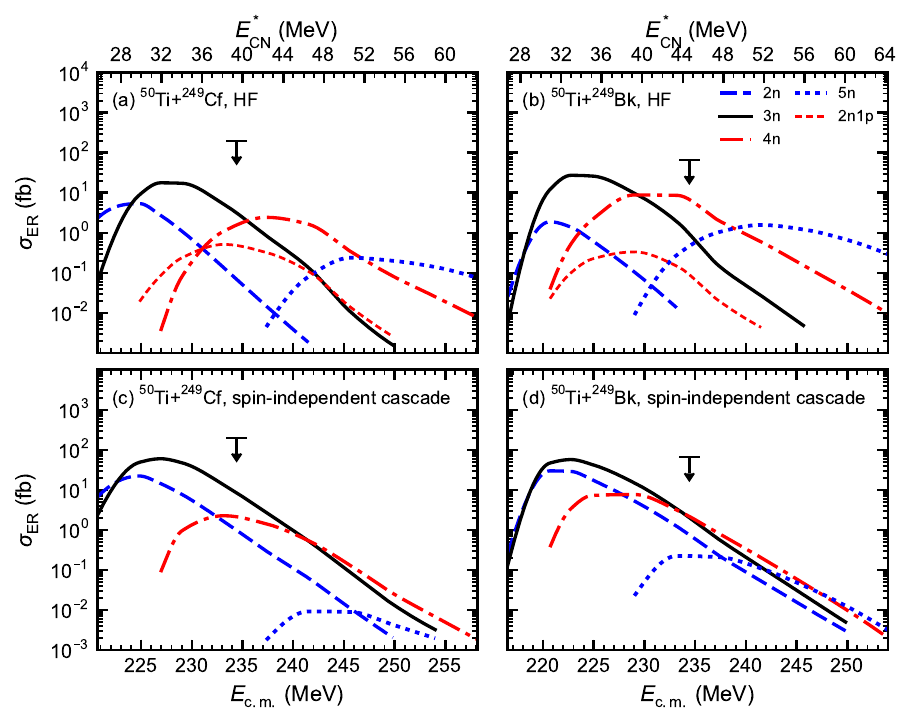}
\caption{Evaporation residue cross sections of the (a), (c) $^{50}$Ti+$^{249}$Cf and (b), (d) $^{50}$Ti+$^{249}$Bk reactions in the 2n--5n and 2n1p channels as functions of $E_{\rm c.m.}$ (bottom axis) and $E^*_{\rm CN}$ (top axis), calculated (a), (b) with the survival probability in the Hauser-Feshbach formalism (Secs.~\ref{sec:cascade}--\ref{sec:levdens}) and (c), (d) by the spin-independent cascade (Sec.~\ref{sec:popcasc}), which includes only neutron emission (2n--5n). The arrows mark the upper limits of the TASCA experiments \cite{Khuyagbaatar2020}.}
\label{fig:er}
\end{figure*}

\begin{table*}[!tbp]
\caption{Capture cross section, fusion probability, and ER cross sections in the 2n--5n and 2n1p channels of the $^{50}$Ti+$^{249}$Cf reaction, obtained with the three definitions of the fusion probability of Sec.~\ref{sec:pcn}: (i) $P_{\rm fus}(t\to\infty)$ of the transport calculation, (ii) Eq.~(\ref{eq:pcn}) with $D_Z(t_0)$ instead of $\tilde D_Z(t_0)$, and (iii) Eq.~(\ref{eq:pcn}) with the normalized $\tilde D_Z(t_0)$ (adopted). The cross sections smaller than 0.1 fb are marked by dashes.}
\label{tab:sigmacf}
{\scriptsize\setlength{\tabcolsep}{1.9pt}\begin{tabular}{rrrrcrrrrrr|rrrrcrrrrrr}
\hline\hline
$E_{\rm lab}$ & $E_{\rm c.m.}$ & $E^*_{\rm CN}$ & $\sigma_{\rm cap}$ & & $\langle P_{\rm CN}\rangle$ & 2n & 3n & 4n & 5n & 2n1p & $E_{\rm lab}$ & $E_{\rm c.m.}$ & $E^*_{\rm CN}$ & $\sigma_{\rm cap}$ & & $\langle P_{\rm CN}\rangle$ & 2n & 3n & 4n & 5n & 2n1p \\
(MeV) & (MeV) & (MeV) & (mb) & & & (fb) & (fb) & (fb) & (fb) & (fb) & (MeV) & (MeV) & (MeV) & (mb) & & & (fb) & (fb) & (fb) & (fb) & (fb) \\
\hline
265 & 220.7 & 25.7 & 18 & (i) & $8.4\times10^{-8}$ & 1.1 & -- & -- & -- & -- & 280 & 233.2 & 38.2 & 117 & (i) & $8.6\times10^{-5}$ & -- & 2.3 & 0.50 & -- & 0.23 \\
 & & & & (ii) & $1.6\times10^{-7}$ & 2.0 & -- & -- & -- & -- &  & & & & (ii) & $1.7\times10^{-4}$ & 0.11 & 4.2 & 0.91 & -- & 0.42 \\
 & & & & (iii) & $1.8\times10^{-7}$ & 2.3 & -- & -- & -- & -- &  & & & & (iii) & $2.1\times10^{-4}$ & 0.13 & 5.1 & 1.1 & -- & 0.51 \\[3pt]
270 & 224.8 & 29.8 & 35 & (i) & $7.2\times10^{-6}$ & 2.5 & 4.1 & -- & -- & -- & 285 & 237.3 & 42.3 & 181 & (i) & $1.4\times10^{-4}$ & -- & 0.38 & 1.1 & -- & 0.14 \\
 & & & & (ii) & $1.4\times10^{-5}$ & 4.6 & 7.7 & -- & -- & -- &  & & & & (ii) & $2.8\times10^{-4}$ & -- & 0.68 & 1.9 & -- & 0.26 \\
 & & & & (iii) & $1.7\times10^{-5}$ & 5.4 & 8.9 & -- & -- & -- &  & & & & (iii) & $3.6\times10^{-4}$ & -- & 0.85 & 2.4 & -- & 0.32 \\[3pt]
272.5 & 226.9 & 31.9 & 48 & (i) & $1.7\times10^{-5}$ & 1.3 & 8.3 & -- & -- & -- & 290 & 241.5 & 46.5 & 235 & (i) & $2.1\times10^{-4}$ & -- & -- & 0.72 & -- & -- \\
 & & & & (ii) & $3.5\times10^{-5}$ & 2.3 & 15 & -- & -- & -- &  & & & & (ii) & $4.1\times10^{-4}$ & -- & 0.11 & 1.3 & -- & -- \\
 & & & & (iii) & $4.1\times10^{-5}$ & 2.7 & 18 & -- & -- & -- &  & & & & (iii) & $5.4\times10^{-4}$ & -- & 0.14 & 1.6 & -- & 0.11 \\[3pt]
275 & 229.0 & 34.0 & 65 & (i) & $3.5\times10^{-5}$ & 0.52 & 8.1 & -- & -- & 0.11 & 295 & 245.7 & 50.6 & 277 & (i) & $2.8\times10^{-4}$ & -- & -- & 0.14 & 0.10 & -- \\
 & & & & (ii) & $7.0\times10^{-5}$ & 0.95 & 15 & -- & -- & 0.20 &  & & & & (ii) & $5.5\times10^{-4}$ & -- & -- & 0.26 & 0.18 & -- \\
 & & & & (iii) & $8.3\times10^{-5}$ & 1.1 & 17 & -- & -- & 0.24 &  & & & & (iii) & $7.4\times10^{-4}$ & -- & -- & 0.34 & 0.24 & -- \\
\hline\hline
\end{tabular}
}

\vspace{10pt}
\caption{The same as in Table~\ref{tab:sigmacf} for the $^{50}$Ti+$^{249}$Bk reaction.}
\label{tab:sigmabk}
{\scriptsize\setlength{\tabcolsep}{1.9pt}\begin{tabular}{rrrrcrrrrrr|rrrrcrrrrrr}
\hline\hline
$E_{\rm lab}$ & $E_{\rm c.m.}$ & $E^*_{\rm CN}$ & $\sigma_{\rm cap}$ & & $\langle P_{\rm CN}\rangle$ & 2n & 3n & 4n & 5n & 2n1p & $E_{\rm lab}$ & $E_{\rm c.m.}$ & $E^*_{\rm CN}$ & $\sigma_{\rm cap}$ & & $\langle P_{\rm CN}\rangle$ & 2n & 3n & 4n & 5n & 2n1p \\
(MeV) & (MeV) & (MeV) & (mb) & & & (fb) & (fb) & (fb) & (fb) & (fb) & (MeV) & (MeV) & (MeV) & (mb) & & & (fb) & (fb) & (fb) & (fb) & (fb) \\
\hline
265 & 220.7 & 30.9 & 21 & (i) & $3.7\times10^{-6}$ & 1.0 & 6.4 & -- & -- & -- & 280 & 233.2 & 43.4 & 187 & (i) & $1.7\times10^{-4}$ & -- & 0.98 & 4.4 & 0.14 & -- \\
 & & & & (ii) & $6.5\times10^{-6}$ & 1.7 & 11 & -- & -- & -- &  & & & & (ii) & $2.9\times10^{-4}$ & -- & 1.6 & 7.3 & 0.23 & 0.13 \\
 & & & & (iii) & $7.0\times10^{-6}$ & 1.9 & 12 & -- & -- & -- &  & & & & (iii) & $3.4\times10^{-4}$ & -- & 1.9 & 8.6 & 0.28 & 0.15 \\[3pt]
267.5 & 222.8 & 33.0 & 34 & (i) & $1.6\times10^{-5}$ & 0.75 & 15 & 0.29 & -- & -- & 285 & 237.3 & 47.6 & 248 & (i) & $2.5\times10^{-4}$ & -- & -- & 1.0 & 0.52 & -- \\
 & & & & (ii) & $2.8\times10^{-5}$ & 1.3 & 25 & 0.49 & -- & -- &  & & & & (ii) & $4.2\times10^{-4}$ & -- & 0.12 & 1.6 & 0.85 & -- \\
 & & & & (iii) & $3.1\times10^{-5}$ & 1.4 & 27 & 0.53 & -- & 0.10 &  & & & & (iii) & $5.2\times10^{-4}$ & -- & 0.15 & 2.0 & 1.0 & -- \\[3pt]
270 & 224.8 & 35.1 & 50 & (i) & $3.9\times10^{-5}$ & 0.35 & 15 & 1.2 & -- & 0.12 & 290 & 241.5 & 51.7 & 308 & (i) & $3.5\times10^{-4}$ & -- & -- & 0.29 & 0.77 & -- \\
 & & & & (ii) & $6.7\times10^{-5}$ & 0.58 & 24 & 1.9 & -- & 0.20 &  & & & & (ii) & $5.9\times10^{-4}$ & -- & -- & 0.48 & 1.3 & -- \\
 & & & & (iii) & $7.4\times10^{-5}$ & 0.64 & 26 & 2.1 & -- & 0.22 &  & & & & (iii) & $7.5\times10^{-4}$ & -- & -- & 0.60 & 1.6 & -- \\[3pt]
275 & 229.0 & 39.2 & 105 & (i) & $1.0\times10^{-4}$ & -- & 5.1 & 4.6 & -- & 0.17 & 295 & 245.7 & 55.9 & 382 & (i) & $4.2\times10^{-4}$ & -- & -- & -- & 0.59 & -- \\
 & & & & (ii) & $1.7\times10^{-4}$ & -- & 8.5 & 7.7 & -- & 0.29 &  & & & & (ii) & $7.0\times10^{-4}$ & -- & -- & 0.12 & 0.95 & -- \\
 & & & & (iii) & $1.9\times10^{-4}$ & 0.11 & 9.8 & 8.8 & -- & 0.33 &  & & & & (iii) & $9.3\times10^{-4}$ & -- & -- & 0.15 & 1.2 & -- \\
\hline\hline
\end{tabular}
}
\end{table*}

The ER cross sections calculated with the HF survival probabilities are presented in Figs.~\ref{fig:er}(a) and \ref{fig:er}(b) and in Tables~\ref{tab:sigmacf} and \ref{tab:sigmabk}. For both reactions the largest cross section is obtained in the 3n channel: 18 fb at $E^*_{\rm CN}=31.9$ MeV for $^{50}$Ti+$^{249}$Cf and 27 fb at 33.0 MeV for $^{50}$Ti+$^{249}$Bk. The 4n excitation functions have their maxima about 10 and 6 MeV higher, with 2.4 and 8.8 fb, respectively, while the 2n, 5n, and 2n1p channels are considerably weaker. The larger cross sections of the $^{50}$Ti+$^{249}$Bk reaction are mainly caused by the larger survival probability of the odd-$Z$ compound nucleus, since the fusion probabilities of the two reactions at the same excitation energy differ by less than a factor of 2 above 33 MeV. Since the three definitions of the fusion probability differ by a factor that depends weakly on the collision energy, the shape of the excitation functions does not depend on the definition; near the maxima, the cross sections obtained with the definitions (ii) and (i) are about 0.9 and 0.5 of the adopted ones for both reactions.

The ER cross sections obtained by the spin-independent cascade are shown in Figs.~\ref{fig:er}(c) and \ref{fig:er}(d). The maxima of the 3n and 4n excitation functions are obtained at the same energies as in the HF calculation, except the 4n maximum of $^{50}$Ti+$^{249}$Cf, which is reached at a lower energy, since the barrier of Eq.~(\ref{eq:bfT}) decreases with the excitation energy faster than the effective barrier of Eq.~(\ref{eq:beff}). The maximum 3n cross sections of the cascade, 60 and 57 fb, are larger than the HF ones by factors of 3.4 and 2.1, while the 4n maxima, 2.3 and 7.7 fb, are similar to them.

\subsection{Comparison with the experimental data}
\label{sec:tasca}

In the TASCA experiments \cite{Khuyagbaatar2020}, no decay chain was observed at $E_{\rm lab}=281.5$ MeV, corresponding to $E^*_{\rm CN}=39.4$ MeV for $^{50}$Ti+$^{249}$Cf and 44.6 MeV for $^{50}$Ti+$^{249}$Bk, at the sensitivity levels of 200 and 65 fb, respectively. At this energy, the calculated sums of the 3n and 4n cross sections, interpolated between $E_{\rm lab}=280$ and 285 MeV, are 4.4 and 6.5 fb, i.e., 45 and 10 times below the experimental limits; with the spin-independent cascade they are 10 and 4 fb. Thus, the present results are consistent with the non-observation of elements 119 and 120 in these experiments. It should be noted that the energy of the experiments lies 7.5 and 11.6 MeV above the maxima of the 3n excitation functions, and that the maximum cross sections are also below the TASCA sensitivity. On the other hand, the cross sections measured for the production of livermorium with $^{50}$Ti, 320 and 220 fb in the 3n and 4n channels of $^{50}$Ti+$^{242}$Pu at $E^*=41$ MeV \cite{Oganessian2025} and 440 fb for $^{50}$Ti+$^{244}$Pu \cite{Gates2024}, are larger than the present results by one order of magnitude, as expected from the decrease of the fusion and survival probabilities with increasing charge of the compound nucleus.

\subsection{Angular momentum window and angular distribution of the evaporation residues}
\label{sec:angle}

\begin{figure*}[!tbp]
\includegraphics[width=0.9\textwidth]{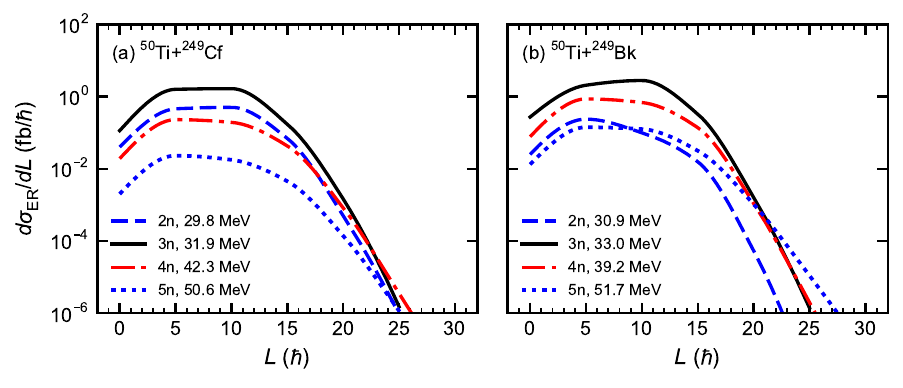}
\caption{Partial ER cross sections $d\sigma_{\rm ER}/dL$ of the (a) $^{50}$Ti+$^{249}$Cf and (b) $^{50}$Ti+$^{249}$Bk reactions for the 2n--5n channels at the maxima of their excitation functions (the excitation energies are given in the legends; grid step $5\hbar$).}
\label{fig:window}
\end{figure*}

The partial ER cross section $\sigma^{xn}_{\rm ER}(E_{\rm c.m.},L)$ is used to define the angular momentum window of the residues as the narrowest interval of $L$ that contains 90\% of $\sigma^{xn}_{\rm ER}$. The partial ER cross sections are shown in Fig.~\ref{fig:window}. For both reactions and all channels the ERs are formed from the partial waves $L\lesssim15\hbar$, with the mean values $\langle L\rangle\approx6$--$8\hbar$ at the maxima of the excitation functions. This window is determined by the decrease of $W_{\rm sur}$ with $L$ [Figs.~\ref{fig:wsur}(c) and \ref{fig:wsur}(d)], while $\sigma_{\rm CN}(L)$ increases up to $L\approx15$--$35\hbar$ at the maxima of the 3n--5n channels. Consequently, only a small part of the fusion cross section contributes to the residues.

In complete fusion the compound nucleus takes over the full momentum of the projectile and moves along the beam direction, independently of the rotation of the DNS before fusion. The direction of the residue is therefore determined by the recoil of the evaporated particles. The angular and energy distributions of the ERs are calculated by Monte Carlo simulation: for every event, $L$ is sampled from $\sigma^{xn}_{\rm ER}(L)$ inside the window, and at every step the thermal energy $U=E^*-E_{\rm rot}(L)$ determines the temperature $T=\sqrt{U/a}$, $a=A/10$ MeV$^{-1}$. The neutron energies are sampled from the Weisskopf spectrum $\varepsilon\exp(-\varepsilon/T)$ with a constant inverse cross section \cite{Weisskopf1937}, the proton energies from the same spectrum above the Coulomb barrier, the emission is isotropic in the frame of the emitter, and the separation energies are taken from FRDM2012 \cite{Moller2016}. This recoil calculation is decoupled from the HF cascade, and its approximations affect only the width of the angular distribution and not the cross sections. The target thickness, the energy loss and multiple scattering in the target, and the acceptance of the separator are not included, so the calculated distributions are the intrinsic ones.

\begin{figure*}[!tbp]
\includegraphics[width=0.9\textwidth]{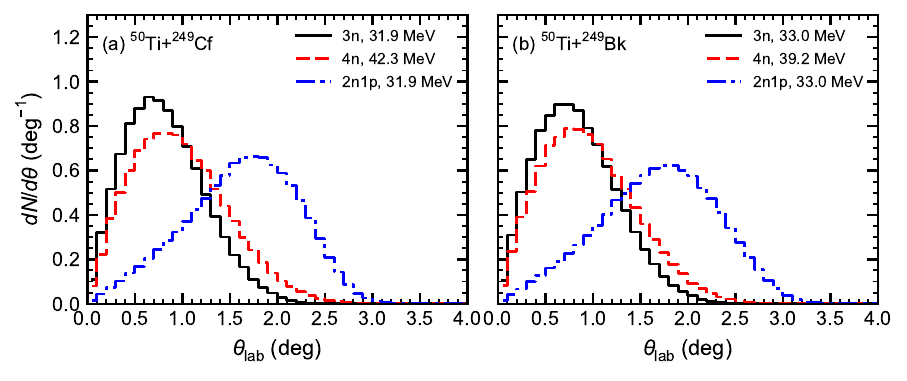}
\caption{Laboratory angular distributions of the evaporation residues of the 3n and 4n channels at the maxima of their excitation functions and of the 2n1p channel at the maximum of the 3n channel for the (a) $^{50}$Ti+$^{249}$Cf and (b) $^{50}$Ti+$^{249}$Bk reactions, without the effects of the target and the separator.}
\label{fig:angle}
\end{figure*}

Figure~\ref{fig:angle} presents the intrinsic laboratory angular distributions of the residues. The xn residues of both reactions are emitted close to the beam direction, with the mean angle of about 0.8$^\circ$ and 90\% of the residues within 1.4$^\circ$ for the 3n channel; the distribution of the 4n residues is slightly broader, and the emission of a proton broadens the distribution to the mean angle of about 1.6--1.7$^\circ$. The mean kinetic energy of the residues of the xn channels is 44--47 MeV.

\subsection{Sensitivity to the fission barrier, the mass table, and the normalization}
\label{sec:senser}

\begin{table*}[!tbp]
\caption{Sensitivity of the ER cross sections to the normalization $N_f$, the mass table, and the fission barrier: maxima of the 3n and 4n cross sections with the excitation energies at which they are reached on the calculated energy grid, and the sums of the 3n and 4n cross sections at the energy of the TASCA experiments, $E_{\rm lab}=281.5$ MeV. For the alternative mass tables and barriers, $N_f$ has been refitted on the $^{48}$Ca+$^{244}$Pu data. One ingredient is changed at a time with respect to the baseline.}
\label{tab:sens}
\begin{tabular}{l|ccccc|ccccc}
\hline\hline
 & \multicolumn{5}{c|}{$^{50}$Ti+$^{249}$Cf} & \multicolumn{5}{c}{$^{50}$Ti+$^{249}$Bk} \\
Variant & $\sigma^{3n}_{\max}$ & $E^*_{\rm CN}$ & $\sigma^{4n}_{\max}$ & $E^*_{\rm CN}$ & $\sigma^{3n+4n}_{\rm TASCA}$ & $\sigma^{3n}_{\max}$ & $E^*_{\rm CN}$ & $\sigma^{4n}_{\max}$ & $E^*_{\rm CN}$ & $\sigma^{3n+4n}_{\rm TASCA}$ \\
 & (fb) & (MeV) & (fb) & (MeV) & (fb) & (fb) & (MeV) & (fb) & (MeV) & (fb) \\
\hline
baseline (FRDM2012, WS+YPE) & 18 & 31.9 & 2.4 & 42.3 & 4.4 & 27 & 33.0 & 8.8 & 39.2 & 6.5 \\
$N_f=0.155$ & 33 & 31.9 & 5.5 & 42.3 & 8.7 & 51 & 33.0 & 20 & 39.2 & 14 \\
$N_f=0.23$ & 10 & 31.9 & 1.1 & 42.3 & 2.3 & 15 & 33.0 & 4.1 & 39.2 & 3.1 \\
KTUY masses & 0.99 & 34.0 & 0.054 & 46.5 & 0.23 & 2.5 & 35.1 & 0.40 & 43.4 & 0.37 \\
WS4 masses & 2.5 & 31.9 & 0.13 & 42.3 & 0.48 & 15 & 33.0 & 1.5 & 39.2 & 1.4 \\
barriers of Ref.~\cite{Moller2015} & 0.68 & 34.0 & 0.019 & 38.2 & 0.14 & 1.2 & 33.0 & 0.13 & 35.1 & 0.031 \\
$\delta W_{\rm sd}=0$ & 24 & 31.9 & 2.3 & 38.2 & 4.9 & 19 & 33.0 & 5.8 & 35.1 & 1.1 \\
\hline\hline
\end{tabular}

\end{table*}

The dependence of the absolute cross sections on the fission barrier and on the ground-state masses is shown in Table~\ref{tab:sens}. The values of $N_f$ for which the calculated cross sections of $^{48}$Ca+$^{244}$Pu are on average larger or smaller than the measured ones by a factor of 2, $N_f=0.155$ and 0.23, change the predicted maxima approximately by the same factor. At fixed $N_f$, the alternative barriers and the WS4 masses give the cross sections of $^{48}$Ca+$^{244}$Pu larger by factors of 30--800, and the refit of $N_f$ removes most of this difference also for the $^{50}$Ti-induced reactions: the 3n maxima change by less than a factor of 1.5 when the saddle-point shell correction is neglected, and they are reduced by factors of 2--7 with the WS4 masses. The largest reductions, by factors of 11--18 with the KTUY masses and about 25 with the barriers of Ref.~\cite{Moller2015}, are caused by the different isotopic dependence of these quantities: the barriers of Ref.~\cite{Moller2015} are higher than the WS+YPE ones by 3.2--3.9 MeV for the flerovium isotopes but by 1.4--2.0 MeV for the isotopes of elements 119 and 120, and the KTUY masses give larger neutron separation energies of the even isotopes along the 3n chains, by up to 0.5 MeV. Thus, the calibration on a reaction with a similar compound nucleus reduces the uncertainty of the predictions considerably, but the differences between the models in the evolution of the shell effects from $Z=114$ to $Z=119$ and 120 are not removed by one normalization factor. For all variants, the calculated sums of the 3n and 4n cross sections at the TASCA energy are below the experimental limits.

\subsection{Comparison with other models}
\label{sec:models}

\begin{table*}[!tbp]
\caption{Maximum ER cross sections in the 3n and 4n channels of the $^{50}$Ti+$^{249}$Cf and $^{50}$Ti+$^{249}$Bk reactions predicted by different models, one row per work, with the excitation energies at which they are reached given in parentheses. SM, statistical model; FBD, fusion-by-diffusion model; DyPES, dynamical potential energy surface; FT-CDFT, finite-temperature covariant density functional theory. The excitation energies of the works that give $E_{\rm c.m.}$ are obtained with the $Q_{gg}$ values of this work. Ranges are given for the calculations with several mass tables \cite{Chen2024,WangFeng2026} or models \cite{FangZhu2026}, and the two values of Ref.~\cite{Denisov2024} correspond to the barriers of FRDM and WS+YPE. Dashes mark the values not given in the original works.}
\label{tab:models}
{\footnotesize\renewcommand{\arraystretch}{0.9}\begin{tabular}{ll|cc|cc}
\hline\hline
 & & \multicolumn{2}{c|}{$^{50}$Ti+$^{249}$Cf} & \multicolumn{2}{c}{$^{50}$Ti+$^{249}$Bk} \\
Model & Ref. & $\sigma_{3n}$ ($E^*_{\rm CN}$) & $\sigma_{4n}$ ($E^*_{\rm CN}$) & $\sigma_{3n}$ ($E^*_{\rm CN}$) & $\sigma_{4n}$ ($E^*_{\rm CN}$) \\
 & & fb (MeV) & fb (MeV) & fb (MeV) & fb (MeV) \\
\hline
Langevin + SM & \cite{Zagrebaev2008} & 42 (38) & 45 (43) & 38 (41) & 56 (43) \\
DNS + SM & \cite{Nasirov2011} & 100 (30) & 4 (36) &  &  \\
SM + empirical $P_{\rm CN}$ & \cite{WangTianScheid2011} & 20 (32) & 4.7 (45) & 35 (33) & 34 (43) \\
DNS (DyPES) & \cite{Wang2012} & 50 (36) & 3.5 (50) & 60 (35) & 110 (45) \\
FBD & \cite{SiwekWilczynska2012} & 6 (34) & 6 (44) & 30 (35) & 30 (42) \\
DNS-type + SM & \cite{Zhu2014} & 20 (34) & 29 (45) & 12 (34) & 64 (46) \\
two-step model & \cite{Liu2016} & 2.7 (37) & 1.6 (46) & 33 & -- \\
proximity + SM & \cite{Santhosh2016} & 103 (31) & 2.3 (42) &  &  \\
DNS & \cite{Adamian2020} & -- & 8 (38) & -- & 19 (40) \\
empirical $P_{\rm CN}$ + SM & \cite{Lv2021} & 7.7 (39) & -- & 48 (36) & 5.7 (54) \\
DNS & \cite{Zargini2023} & 51 (33) & 1.1 (36) & 8 (36) & 3.3 (38) \\
semianalytic (FRDM / WS+YPE $B_f$) & \cite{Denisov2024} & 0.22 / 0.12 (35 / 38) & 0.31 / 0.022 (38 / 48) & 8.5 / 2.0 (31 / 33) & 7.0 / 0.7 (37 / 43) \\
DNS, five mass tables & \cite{Chen2024} & 0.6--8.8 (34--40) & -- & 0.6--13 (32--36) & -- \\
DNS & \cite{ZhangMH2024} & 2.3 (37) & 0.3 (46) & 8 (34) & 10 (40) \\
DNS + SM & \cite{Qiang2024} &  &  & 220 (35) & -- \\
DNS (Bayesian) & \cite{Fang2024} &  &  & $1.2\times10^3$ (33) & 120 (40) \\
DNS + SM & \cite{ZhuZhaoBao2024} &  &  & 17 (37) & 6 (40) \\
DNS & \cite{Yao2025} & 1.6 (36) & -- & 27 (35) & -- \\
DNS, five mass tables & \cite{WangFeng2026} & 16--940 (31--33) & 0.3--40 (40--42) & 7.9--650 (30--31) & 8.4--580 (39--40) \\
DNS & \cite{Li2026} & 1.8 (38) & -- & 22 (36) & 4.7 (44) \\
DNS (FT-CDFT inputs) & \cite{Zhang2026} & 13 (34) & 48 (41) &  &  \\
diffusion models & \cite{FangZhu2026} &  &  & 85--110 (33) & 18--56 (40--42) \\
Langevin + SM & \cite{Kawai2026} &  &  & 52 (40) & 145 (41) \\
semiempirical & \cite{WangDuanYao2026} &  &  & -- & 13 (39) \\
\hline
DNS + HF & This work & 18 (31.9) & 2.4 (42.3) & 27 (33.0) & 8.8 (39.2) \\
DNS + spin-independent cascade & This work & 60 (31.9) & 2.3 (38.2) & 57 (33.0) & 7.7 (39.2) \\
\hline\hline
\end{tabular}
}
\end{table*}

The present results are compared with the predictions of other models in Table~\ref{tab:models}. The predicted maxima of the 3n and 4n cross sections span about four orders of magnitude, and most of them lie between a few fb and 100 fb. The present 3n cross sections are close to the median values of the predictions for both reactions, and the present ratio of the 4n and 3n cross sections is within the range of the other models. The dispersion of the theoretical results is related, first of all, to the different fission barriers and ground-state masses used in the calculations \cite{SiwekWilczynska2012}; this can be seen, for example, in the results of the same DNS calculations with different mass tables \cite{Chen2024,WangFeng2026}, which differ by one to two orders of magnitude, and in the sensitivity study of Sec.~\ref{sec:senser}.

\section{DECAY CHAINS OF THE EVAPORATION RESIDUES}
\label{sec:chains}

The decay of the ER nuclei is followed through the sequence of $\alpha$ decays until spontaneous fission (SF) dominates. The $Q_\alpha$ values are taken from the experimental data where available: AME2020 \cite{AME2020b} and the measurements published after this evaluation, $^{288}$Lv and $^{289}$Lv \cite{Oganessian2025} and $^{275}$Ds \cite{Oganessian2024}. For the unknown nuclei the WS4 mass model \cite{Wang2014} is used, while the model uncertainty is estimated with FRDM2012 \cite{Moller2016}. For the 65 $\alpha$ emitters with $Z\ge104$ and measured $Q_\alpha$, the root-mean-square deviation of the WS4 values from experiment is 0.24 MeV, while FRDM2012 gives 0.52 MeV.

The $\alpha$-decay half-lives are calculated with three formulas: the Viola-Seaborg relation with the parameters of Sobiczewski \textit{et al.} \cite{Viola1966,Sobiczewski1989},
\begin{equation}
\log_{10}T_\alpha({\rm s})=\frac{aZ+b}{\sqrt{Q_\alpha}}+cZ+d+h_{\log},
\label{eq:vss}
\end{equation}
with $a=1.66175$, $b=-8.5166$, $c=-0.20228$, $d=-33.9069$, and $h_{\log}=0$, 0.772, 1.066, and 1.114 for the even-even, odd-$Z$, odd-$N$, and odd-odd nuclei, respectively; the formula of Royer \cite{Royer2000} with its separate parameter sets for the four parity classes; and the universal decay law of Qi \textit{et al.} \cite{Qi2009}, which was fitted to even-even nuclei and gives a lower limit for odd nuclei. The $Q_\alpha$ values of the ground-state-to-ground-state transitions are used, although odd nuclei often decay to excited states, which increases their half-lives. For the 65 nuclei with $Z\ge104$ and known half-lives, these formulas reproduce the measured partial $\alpha$-decay half-lives with the root-mean-square deviations of $\log_{10}T_\alpha$ equal to 0.75, 0.82, and 1.23, respectively. The central value is the median of the three formulas; the uncertainty range includes the spread over the formulas and over the $Q_\alpha$ values of WS4 and FRDM2012.

The SF half-lives are calculated with the formulas of Xu \textit{et al.} \cite{Xu2008}, Ren and Xu \cite{RenXu2005}, and Santhosh \textit{et al.} \cite{Santhosh2010}. The first and the last were fitted to even-even nuclei; for odd nuclei the average of $\log_{10}T_{\rm SF}$ of the neighboring even-even nuclei is used. Since all three formulas underestimate the half-lives of odd nuclei by 3--4 orders of magnitude, each SF formula is corrected by its average deviation from the data for the given parity class, determined from 15 even-even, 16 odd-$N$, 6 odd-$Z$, and 5 odd-odd nuclei with $Z\ge104$; the corrections for the last two classes are therefore uncertain. The decay chain is followed while the $\alpha$ branch of the nucleus exceeds 1\%, and the measured half-lives and branchings are used for the known members of the chain. The $\beta^+$ decay and electron capture are not considered; they may compete with SF for the long-lived end members of the chains, but not with the fast $\alpha$ decays of the members with $Z\ge110$, which are used for the identification.

\begin{figure*}[!tbp]
\includegraphics[width=0.9\textwidth]{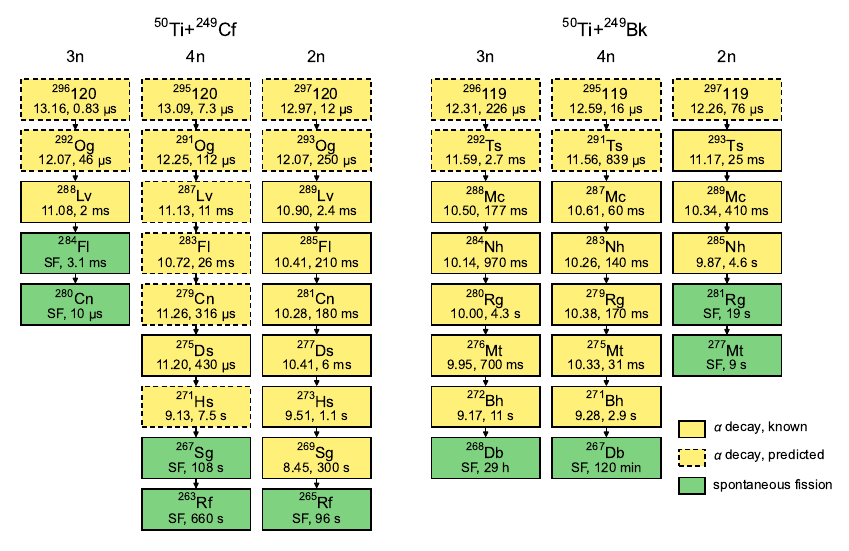}
\caption{Predicted decay chains of the residues of the 3n, 4n, and 2n channels of the $^{50}$Ti+$^{249}$Cf and $^{50}$Ti+$^{249}$Bk reactions. Each box gives the nucleus, the $\alpha$-particle energy $E_\alpha$ (MeV) or the spontaneous fission (SF), and the half-life. Solid boxes: known nuclei with the measured half-lives \cite{NUBASE2020,Oganessian2025,Oganessian2024}; dashed boxes: nuclei not observed so far, with $Q_\alpha$ from WS4 and the median $T_\alpha$ of three formulas. Yellow boxes decay by $\alpha$ emission and green boxes by SF. The uncertainty ranges of the half-lives, the SF half-lives, and the $Q_\alpha$ values of Ref.~\cite{Adamian2020} are given in Table~\ref{tab:chainsfull}.}
\label{fig:chains}
\end{figure*}

The predicted decay chains are shown in Fig.~\ref{fig:chains}, and the full table with the uncertainty ranges of the half-lives is given in Table~\ref{tab:chainsfull}. The isotopes of element 120 are expected to decay by the emission of $\alpha$ particles with the energies $E_\alpha=13.0$--13.2 MeV and the half-lives of about 1 $\mu$s for the even-even $^{296}$120 (3n) and 7--12 $\mu$s for the odd $^{295,297}$120 (4n, 2n). The 3n chain $^{296}$120$\to^{292}$Og$\to^{288}$Lv$\to^{284}$Fl reaches the recently discovered $^{288}$Lv \cite{Oganessian2025}, and the 2n chain reaches $^{289}$Lv and continues through the known nuclei $^{285}$Fl, $^{281}$Cn, $^{277}$Ds, $^{273}$Hs, and $^{269}$Sg to the spontaneous fission of $^{265}$Rf.

The isotopes of element 119 are expected to decay by the emission of $\alpha$ particles with $E_\alpha=12.3$--12.6 MeV and longer half-lives, 16--230 $\mu$s, mainly because of their smaller $Q_\alpha$ values. The 3n chain $^{296}$119$\to^{292}$Ts passes through one unknown tennessine isotope and continues through the known nuclei $^{288}$Mc, $^{284}$Nh, $^{280}$Rg, $^{276}$Mt, and $^{272}$Bh to $^{268}$Db, which decays by SF or electron capture; the 4n chain $^{295}$119$\to^{291}$Ts reaches $^{287}$Mc and continues through known nuclei to $^{267}$Db, and the 2n chain passes through the known $^{293}$Ts. These long chains of $\alpha$ decays ending in well-known nuclei would provide a reliable identification of the new elements.

The $Q_\alpha$ values of the parent nuclei $^{295}$119 and $^{295,297}$120 predicted in Ref.~\cite{Adamian2020} with the single-particle potential extracted from the self-consistent mean-field calculations are smaller than the WS4 values by 0.9--1.6 MeV, since that potential gives a stronger shell effect at $Z=120$. In that work, the ground-state transition of $^{295}$120 is hindered, and its $\alpha$ decay is expected in $T_\alpha\approx20$ ms with $Q_\alpha=11.58$ MeV; the chains of $^{295}$119, $^{295}$120, and $^{297}$120 are predicted to be terminated by spontaneous fission below $^{279}$Rg, $^{279}$Cn, and $^{285}$Fl, respectively, although the nuclei $^{275}$Mt, $^{275}$Ds, and $^{281}$Cn that follow in these chains are known $\alpha$ emitters \cite{NUBASE2020,Oganessian2024}. For the daughter nuclei the two predictions agree within 0.7 MeV. Thus, the half-lives of the parent nuclei are the most uncertain part of the predicted chains and may range from microseconds to tens of milliseconds.

It should be stressed that the predicted half-lives of the isotopes of element 120 are comparable with the flight time of the residues through a recoil separator, which is of the order of 1 $\mu$s: the residues with the kinetic energies of about 45 MeV obtained in Sec.~\ref{sec:angle} move with the velocity of about 0.018$c$ and pass a separator of 3 m length in about 0.5 $\mu$s. Some of the $^{296}$120 nuclei can decay before implantation, and the detection of the $\alpha$ decay of the implanted nuclei requires pulse processing that can resolve signals separated by less than 1 $\mu$s. This should be taken into account in the estimate of the efficiency of the experiments.

\section{CONCLUSIONS}
\label{sec:concl}

The formation of elements 119 and 120 in the $^{50}$Ti+$^{249}$Bk and $^{50}$Ti+$^{249}$Cf reactions has been studied in the DNS model, with the fusion probability calculated from the charge distribution of the DNS obtained with detailed-balance transition rates, and with the survival probability calculated in the HF formalism for every excitation energy and angular momentum of the compound nucleus. The fusion probability exceeds the probability of the diffusion over the maximum of the driving potential by a factor of 2--3, and above $E^*_{\rm CN}\approx35$ MeV it is similar for the two reactions at the same excitation energy. In the survival calculation, the shell correction is damped once, through the level densities at the ground state and at the triaxial saddle points of the macroscopic-microscopic barriers, and the normalization of the saddle-point level density has been fixed by the measured cross sections of the $^{48}$Ca+$^{244}$Pu reaction. The largest ER cross sections are obtained in the 3n channel, 18 fb at $E^*_{\rm CN}=32$ MeV for $^{50}$Ti+$^{249}$Cf and 27 fb at 33 MeV for $^{50}$Ti+$^{249}$Bk; the 4n maxima, 2.4 and 9 fb, are reached at 42 and 39 MeV. At the energy of the TASCA experiments, the calculated sums of the 3n and 4n cross sections, 4 and 6 fb, are well below the experimental limits, which is consistent with the non-observation of these elements. With the normalization refitted, the alternative mass tables and fission barriers considered here give the maximum 3n cross sections smaller by up to a factor of 25, and the sums at the TASCA energy remain below the limits. The residues are formed in the partial waves $L\lesssim15\hbar$, and the evaporation recoil gives them an intrinsic angular spread of about $1.5^\circ$ around the beam direction. The predicted decay chains show that the isotopes of element 120 should have half-lives of 1--12 $\mu$s and those of element 119 of 16--230 $\mu$s, and that the 3n and 2n chains of element 120 pass through the known isotopes of livermorium, while all chains of element 119 reach the known isotopes of moscovium. These results may serve as a theoretical benchmark for the planned experiments on the synthesis of elements 119 and 120 at the leading research facilities.

\begin{acknowledgments}
The author is grateful to A. K. Nasirov for valuable discussions. During the preparation of this work the author used Claude (Anthropic; Claude Opus 5.5 and Claude Fable 5.1, via Claude Code) to analyze the results of the calculations, to write and test the Python codes used for the data processing and the figures, to check the bibliography, to perform an internal review of the manuscript, and to draft and edit the text under the author's direction. All numerical results and statements were reviewed and verified by the author, who takes full responsibility for the content of the manuscript.
\end{acknowledgments}

\section*{DATA AVAILABILITY}
The calculated cross sections that support the findings of this article are presented in Tables~\ref{tab:sigmacf} and \ref{tab:sigmabk}. Other data are not publicly available and are available from the author upon reasonable request.

\begin{table*}[!tbp]
\caption{Predicted decay chains of the residues of the 3n, 4n, and 2n channels of the $^{50}$Ti+$^{249}$Cf and $^{50}$Ti+$^{249}$Bk reactions. $Q_\alpha$ is experimental where known, otherwise from WS4 (marked); $T_\alpha$ is the median of three formulas with the range over formulas and mass models; $T_{\rm SF}$ is the median of three parity-corrected formulas; the last column gives the measured half-lives.}
\label{tab:chainsfull}
{\footnotesize\begin{tabular*}{\textwidth}{@{\extracolsep{\fill}}lllllll}
\hline\hline
Nucleus & $Q_\alpha$ (MeV) & $Q_\alpha$ \cite{Adamian2020} & $E_\alpha$ (MeV) & $T_\alpha$ & $T_{\rm SF}$ & Measured \\
\hline
\multicolumn{7}{l}{$^{50}$Ti+$^{249}$Cf, 3n channel} \\
$^{296}$120 & 13.34$^{\rm WS4}$ & -- & 13.16 & 0.83 $\mu$s [0.22 $\mu$s--1.2 $\mu$s] & $10^{8}$ y & -- \\
$^{292}$Og & 12.24$^{\rm WS4}$ & -- & 12.07 & 46 $\mu$s [21 $\mu$s--69 $\mu$s] & 105 y & -- \\
$^{288}$Lv & 11.24 & -- & 11.08 & 3.1 ms [2.5 ms--3.9 ms] & 20 h & 2 ms \\
$^{284}$Fl & 10.72 & -- & 10.57 & 16 ms [12 ms--19 ms] & 29 s & 3.1 ms (SF) \\
$^{280}$Cn & 10.86$^{\rm WS4}$ & -- & 10.71 & 1.4 ms [377 $\mu$s--2.1 ms] & 167 ms & 10 $\mu$s (SF) \\
\multicolumn{7}{l}{$^{50}$Ti+$^{249}$Cf, 4n channel} \\
$^{295}$120 & 13.27$^{\rm WS4}$ & 11.99 & 13.09 & 7.3 $\mu$s [0.41 $\mu$s--19 $\mu$s] & $10^{12}$ y & -- \\
$^{291}$Og & 12.42$^{\rm WS4}$ & 11.76 & 12.25 & 112 $\mu$s [9.6 $\mu$s--326 $\mu$s] & $10^{6}$ y & -- \\
$^{287}$Lv & 11.28$^{\rm WS4}$ & 11.35 & 11.13 & 11 ms [2.4 ms--58 ms] & 15 y & -- \\
$^{283}$Fl & 10.88$^{\rm WS4}$ & 11.05 & 10.72 & 26 ms [6.2 ms--66 s] & 53 h & -- \\
$^{279}$Cn & 11.42$^{\rm WS4}$ & 11.09 & 11.26 & 316 $\mu$s [16 $\mu$s--1.1 ms] & 19 min & -- \\
$^{275}$Ds & 11.37 & -- & 11.20 & 104 $\mu$s [18 $\mu$s--401 $\mu$s] & 80 s & 430 $\mu$s \\
$^{271}$Hs & 9.26$^{\rm WS4}$ & -- & 9.13 & 7.5 s [2.6 s--622 s] & 56 s & -- \\
$^{267}$Sg & 8.14$^{\rm WS4}$ & -- & 8.02 & 119 min [62 min--51 h] & 319 s & 108 s (SF) \\
$^{263}$Rf & 8.05$^{\rm WS4}$ & -- & 7.93 & 42 min [21 min--57 h] & 205 min & 660 s (SF) \\
\multicolumn{7}{l}{$^{50}$Ti+$^{249}$Cf, 2n channel} \\
$^{297}$120 & 13.14$^{\rm WS4}$ & 11.52 & 12.97 & 12 $\mu$s [0.16 $\mu$s--35 $\mu$s] & $10^{6}$ y & -- \\
$^{293}$Og & 12.24$^{\rm WS4}$ & 11.54 & 12.07 & 250 $\mu$s [27 $\mu$s--793 $\mu$s] & 6.2 y & -- \\
$^{289}$Lv & 11.05 & -- & 10.90 & 37 ms [9.1 ms--130 ms] & 819 min & 2.4 ms \\
$^{285}$Fl & 10.56 & -- & 10.41 & 160 ms [43 ms--595 ms] & 369 s & 210 ms \\
$^{281}$Cn & 10.43 & -- & 10.28 & 79 ms [20 ms--313 ms] & 16 s & 180 ms \\
$^{277}$Ds & 10.56$^{\rm WS4}$ & -- & 10.41 & 8.2 ms [1.8 ms--34 ms] & 3.7 s & 6 ms \\
$^{273}$Hs & 9.65 & -- & 9.51 & 507 ms [149 ms--2.2 s] & 5.3 s & 1.1 s \\
$^{269}$Sg & 8.58 & -- & 8.45 & 201 s [85 s--899 s] & 28 s & 300 s \\
$^{265}$Rf & 7.42$^{\rm WS4}$ & -- & 7.31 & 168 h [115 h--2.7 y] & 17 min & 96 s (SF) \\
\multicolumn{7}{l}{$^{50}$Ti+$^{249}$Bk, 3n channel} \\
$^{296}$119 & 12.48$^{\rm WS4}$ & -- & 12.31 & 226 $\mu$s [1.9 $\mu$s--515 $\mu$s] & 201 y & -- \\
$^{292}$Ts & 11.75$^{\rm WS4}$ & -- & 11.59 & 2.7 ms [302 $\mu$s--7.6 ms] & 936 h & -- \\
$^{288}$Mc & 10.65 & -- & 10.50 & 449 ms [50 ms--783 ms] & 175 min & 177 ms \\
$^{284}$Nh & 10.28 & -- & 10.14 & 997 ms [116 ms--1.8 s] & 186 s & 970 ms \\
$^{280}$Rg & 10.15 & -- & 10.00 & 456 ms [56 ms--957 ms] & 18 s & 4.3 s \\
$^{276}$Mt & 10.10 & -- & 9.95 & 119 ms [15 ms--297 ms] & 7.2 s & 700 ms \\
$^{272}$Bh & 9.30 & -- & 9.17 & 5.6 s [733 ms--12 s] & 13 s & 11 s \\
$^{268}$Db & 7.85$^{\rm WS4}$ & -- & 7.73 & 41 h [302 min--323 d] & 175 s & 29 h (SF) \\
\multicolumn{7}{l}{$^{50}$Ti+$^{249}$Bk, 4n channel} \\
$^{295}$119 & 12.76$^{\rm WS4}$ & 11.84 & 12.59 & 16 $\mu$s [2.4 $\mu$s--59 $\mu$s] & $10^{4}$ y & -- \\
$^{291}$Ts & 11.72$^{\rm WS4}$ & 11.81 & 11.56 & 839 $\mu$s [327 $\mu$s--3.2 ms] & 953 h & -- \\
$^{287}$Mc & 10.76 & -- & 10.61 & 47 ms [26 ms--184 ms] & 163 min & 60 ms \\
$^{283}$Nh & 10.41$^{\rm WS4}$ & -- & 10.26 & 93 ms [52 ms--202 s] & 173 s & 140 ms \\
$^{279}$Rg & 10.53 & -- & 10.38 & 10 ms [4.9 ms--43 ms] & 17 s & 170 ms \\
$^{275}$Mt & 10.48 & -- & 10.33 & 3.2 ms [1.4 ms--14 ms] & 9.1 s & 31 ms \\
$^{271}$Bh & 9.42 & -- & 9.28 & 579 ms [326 ms--2.4 s] & 25 s & 2.9 s \\
$^{267}$Db & 7.59$^{\rm WS4}$ & -- & 7.48 & 60 h [54 h--167 d] & 333 s & 120 min (SF) \\
\multicolumn{7}{l}{$^{50}$Ti+$^{249}$Bk, 2n channel} \\
$^{297}$119 & 12.42$^{\rm WS4}$ & -- & 12.26 & 76 $\mu$s [2.8 $\mu$s--302 $\mu$s] & 12 y & -- \\
$^{293}$Ts & 11.32 & -- & 11.17 & 6.9 ms [3.6 ms--28 ms] & 61 h & 25 ms \\
$^{289}$Mc & 10.49 & -- & 10.34 & 223 ms [137 ms--948 ms] & 695 s & 410 ms \\
$^{285}$Nh & 10.01 & -- & 9.87 & 1.1 s [699 ms--4.6 s] & 12 s & 4.6 s \\
$^{281}$Rg & 9.76$^{\rm WS4}$ & -- & 9.62 & 1.2 s [784 ms--234 s] & 1.2 s & 19 s (SF) \\
$^{277}$Mt & 9.64$^{\rm WS4}$ & -- & 9.50 & 581 ms [153 ms--2.6 s] & 231 ms & 9 s (SF) \\
\hline\hline
\end{tabular*}
}
\end{table*}

\appendix*
\section{PREDICTED DECAY CHAINS}
\label{sec:appchains}

Table~\ref{tab:chainsfull} gives the predicted decay chains of the evaporation residues of the 3n, 4n, and 2n channels shown in Fig.~\ref{fig:chains}. The $Q_\alpha$ values are experimental where known \cite{AME2020b,Oganessian2025,Oganessian2024}, otherwise from the WS4 mass model \cite{Wang2014}; for comparison, the $Q_\alpha$ values of Ref.~\cite{Adamian2020} are also given. The $\alpha$-decay half-life $T_\alpha$ is the median of the three formulas of Sec.~\ref{sec:chains}, with the range over the formulas and over the $Q_\alpha$ values of WS4 and FRDM2012 in brackets, and $T_{\rm SF}$ is the median of the three parity-corrected formulas. The last column gives the measured half-lives and the decay mode of the known nuclei \cite{NUBASE2020,Oganessian2025,Oganessian2024}.

\bibliography{refs}

\end{document}